\documentclass[twocolumn,floatfix,showpacs,prd,aps,tightenlines]{revtex4}
\usepackage{graphicx}
\usepackage{amsmath}
\usepackage{psfrag}
\usepackage{dcolumn}
\usepackage{bm}

\newcommand{\bea}{\begin{eqnarray}}
\newcommand{\eea}{\end{eqnarray}}
\newcommand{\beq}{\begin{equation}}
\newcommand{\eeq}{\end{equation}}
\newcommand{\KMS}{\rm km\,s^{-1}}
\newcommand{\vz}{v_\|}
\newcommand{\avg}[1]{\langle #1 \rangle}

\begin{document}

\title{Further insight into gravitational recoil}

\author{Carlos O. Lousto}
\affiliation{Center for Computational Relativity and Gravitation,
School of Mathematical Sciences,
Rochester Institute of Technology, 78 Lomb Memorial Drive, Rochester,
 New York 14623}

\author{Yosef Zlochower} 
\affiliation{Center for Computational Relativity and Gravitation,
School of Mathematical Sciences,
Rochester Institute of Technology, 78 Lomb Memorial Drive, Rochester,
 New York 14623}

\date{\today}

\begin{abstract}
We test the accuracy of our recently proposed empirical formula to
model the recoil velocity imparted to the merger remnant of spinning,
unequal-mass black-hole binaries.  We study three families of
black-hole binary configurations,  all with mass ratio q=3/8 (to
maximize the unequal-mass contribution to the kick) and spins aligned
(or counter aligned) with the orbital angular momentum, two with
spin configurations chosen to minimize the spin-induced tangential and
radial accelerations of the trajectories respectively, and a third
family where the trajectories are significantly altered by spin-orbit
coupling. We find good agreement between the measured and predicted recoil
velocities for the first two families, and
reasonable agreement for the third. We also re-examine our original
generic binary configuration that led to the discovery of extremely
large spin-driven recoil velocities and inspired our empirical formula,
and find reasonable agreement between the predicted and measured
recoil speeds.

\end{abstract}

 \maketitle

\section{Introduction}

Thanks to recent breakthroughs in the full non-linear numerical
evolution of black-hole-binary
spacetimes~\cite{Pretorius:2005gq,Campanelli:2005dd,Baker:2005vv}, it
is now possible to accurately simulate the merger process and examine
its effects in this highly non-linear
regime~\cite{Campanelli:2006gf,
Baker:2006yw,Campanelli:2006uy,Campanelli:2006fg,Campanelli:2006fy,
Pretorius:2006tp,Pretorius:2007jn,Baker:2006ha,Bruegmann:2006at,
Buonanno:2006ui,Baker:2006kr,Scheel:2006gg,Baker:2007fb,Marronetti:2007ya,
Pfeiffer:2007yz}.  Black-hole binaries will radiate between $2\%$ and
$8\%$ of their total mass and up to $40\%$ of their angular momenta,
depending on the magnitude and direction of the  spin components,
during the
merger~\cite{Campanelli:2006uy,Campanelli:2006fg,Campanelli:2006fy}.
In addition, the radiation of net linear momentum by a black-hole
binary leads to the recoil of the final remnant
hole~\cite{Campanelli:2004zw, Herrmann:2006ks,
Baker:2006vn,Sopuerta:2006wj,Gonzalez:2006md,
Sopuerta:2006et,Herrmann:2006cd,Herrmann:2007zz,Herrmann:2007ac,
Campanelli:2007ew,Koppitz:2007ev,Choi:2007eu,Gonzalez:2007hi,
Baker:2007gi,Campanelli:2007cga,Berti:2007fi,Tichy:2007hk,
Herrmann:2007ex,Brugmann:2007zj,Schnittman:2007ij,Krishnan:2007pu,
HolleyBockelmann:2007eh,Pollney:2007ss},
which can have astrophysically important effects
\cite{Redmount:1989,Merritt:2004xa,Campanelli:2007ew,Gualandris:2007nm,HolleyBockelmann:2007eh,Kapoor76}.

Merging black-hole binaries will radiate net linear momentum if the two
black holes are not symmetric. This asymmetry can be due to unequal
masses, unequal spins, or a combination of the two.  A non-spinning
black-hole binary will thus only radiate net linear momentum if the
component  masses are not equal.  However, the maximum recoil in this
case (which occurs when the mass ratio is $q\approx0.36$) is a
relatively small $\sim175\ \KMS$~\cite{Gonzalez:2006md}.  The
complementary case, where the black holes have equal masses but
unequal spins was first reported in~\cite{Herrmann:2007ac} and
\cite{Koppitz:2007ev}. In the former case the authors calculated the
recoil velocity for equal-mass, quasi-circular binaries with
equal-amplitude, anti-parallel spins aligned with the orbital angular
momentum direction, while in the latter case the authors used the same
general configuration but varied the
amplitude of one of the spins.  In both the above cases the authors
extrapolated a maximum possible recoil (which is tangent to the
orbital plane) of $~\sim 460\ \KMS$ when the two holes have maximal
spin.  At the same time, our group
released a paper on the first simulation of a generic black-hole
binaries with unequal masses and spins, where the spins were not
aligned with the orbital angular momentum~\cite{Campanelli:2007ew}. That
configuration had a mass ratio of 1:2, with the larger black hole
having spin $a/m = 0.885$ pointing $45^\circ$ below the orbital plane
and the smaller hole having negligible spin.  The black holes displayed spin
precession and spin flips and a measured recoil velocity of $475\
\KMS$, mostly along the orbital angular momentum direction. 
It was thus found that the recoil normal to the orbital plane (due to spin
components lying in the orbital plane) can be larger than the in-plane
recoil originating from either the unequal-masses or the spin components
normal to the orbital plane. The maximum possible recoil arises from
equal-mass, maximally spinning holes with spins in the orbital plane and
counter-aligned. This maximum recoil, which will be normal to the orbital
plane, is nearly $4000\ \KMS$.

In~\cite{Campanelli:2007ew} we introduced the following 
heuristic model for the
gravitational recoil of a merging binary.
\begin{equation}\label{eq:empirical}
\vec{V}_{\rm recoil}(q,\vec\alpha_i)=v_m\,\hat{e}_1+
v_\perp(\cos(\xi)\,\hat{e}_1+\sin(\xi)\,\hat{e}_2)+\vz\,\hat{e}_z,
\end{equation}
where
\begin{subequations}
\begin{equation}\label{eq:vm}
v_m=A\frac{q^2(1-q)}{(1+q)^5}\left(1+B\,\frac{q}{(1+q)^2}\right),
\end{equation}
\begin{equation}\label{eq:vperp}
v_\perp=H\frac{q^2}{(1+q)^5}\left(\alpha_2^\|-q\alpha_1^\|\right),
\end{equation}
\begin{equation}\label{eq:vpar}
\vz=K\cos(\Theta-\Theta_0)\frac{q^2}{(1+q)^5}\left|\vec\alpha_2^\perp-q\vec\alpha_1^\perp\right|,
\end{equation}
\end{subequations}
$A = 1.2\times 10^{4}\ \KMS$~\cite{Gonzalez:2006md},
$B = -0.93$~\cite{Gonzalez:2006md},
here we find $H = (6.9\pm0.5)\times 10^{3}\ \KMS$,
$\vec{\alpha}_i=\vec{S}_i/m_i^2$,
$\vec S_i$ and $m_i$ are the spin and mass of
hole $i$, $q=m_1/m_2$ is the mass ratio of the smaller to larger mass hole,
the index $\perp$ and $\|$
refer to perpendicular and parallel to the orbital angular momentum
respectively at the effective moment of the maximum generation of
the recoil (around merger time),
$\hat{e}_1,\hat{e}_2$ are orthogonal unit vectors in the
orbital plane, and $\xi$ measures the angle between the ``unequal mass''
and ``spin'' contributions to the recoil velocity in the orbital plane.
The angle $\Theta$ was defined as the angle between the in-plane
component of $\vec \Delta\equiv (m_1+m_2)({\vec S_2}/m_2 -{\vec S_1}/m_1)$ 
and the infall direction at merger.
The form of Eq.~(\ref{eq:vm}) was proposed 
in~\cite{1983MNRAS.203.1049F,Gonzalez:2006md},
while the form of Eqs.~(\ref{eq:vperp})~and~(\ref{eq:vpar}) was 
proposed in~\cite{Campanelli:2007ew}
based on the post-Newtonian expressions in~\cite{Kidder:1995zr}.
 In Ref~\cite{Campanelli:2007cg} we 
determined that
$K=(6.0\pm0.1)\times 10^4\ \KMS$.
Although $\xi$ may in general depend strongly on the configuration,
the results of~\cite{Choi:2007eu} and post-Newtonian calculations show that $\xi$ is
$90^\circ$ for headon collisions, and the results presented here
indicate that $\xi \sim 145^\circ$ for a wide range of quasi-circular
configurations.
A simplified version of Eq.~(\ref{eq:empirical}) that
models the magnitude of $V_{\rm recoil}$ was independently proposed
in~\cite{Baker:2007gi}, and a simplified form of Eq.~(\ref{eq:empirical})
for the equal-mass aligned spin case was proposed in~\cite{Koppitz:2007ev}.

Our heuristic formula~(\ref{eq:empirical}) describing the recoil
velocity of a black-hole binary remnant as a function of the
parameters of the individual holes has been theoretically verified in
several ways. In~\cite{Campanelli:2007cg} the $\cos{\Theta}$ dependence
was established and was confirmed in~\cite{Brugmann:2007zj} for 
binaries with larger initial separations. In Ref.~\cite{Herrmann:2007ex}
the decomposition into spin
components perpendicular and parallel to the orbital plane was
verified, and in~\cite{Pollney:2007ss} it was found that the quadratic-in-spin
corrections to the in-plane recoil velocity are less than
$20\ \KMS$.

Consistent and independent recoil velocity
calculations have also been obtained for equal-mass binaries with
spinning black holes that have spins aligned/counter-aligned with the
orbital angular momentum~\cite{Herrmann:2007ac,Koppitz:2007ev}. Recoils
from the merger of non-precessing unequal mass black-hole binaries have been
modeled in~\cite{Baker:2007gi}.

The net in-plane remnant recoil velocity arises both from the 
asymmetry due to unequal masses, which given its $z\to-z$ symmetric
behavior, only contributes to recoil along the orbital plane, and
the asymmetry produced by the black-hole spin
component perpendicular to the orbital plane. Even if we can parametrize
the contribution of each of these two components of the recoil in terms
of only one angle, $\xi$, the modeling of it appears in principle very
complicated. $\xi$ may depend on the mass ratio ($q$) of the holes,
as well as their individual spins $S^z_1$ and $S^z_2$, but also on their
orbital parameters such as initial coordinates and momenta,
or initial separation and eccentricity. We clearly have to reduce
the dimensionality of this parameter space as part of the modeling process.
In order to do so, we shall choose a model for $\xi$ that only depends on
$q$ and $\Delta^z$  for quasi-circular orbits. We then perform simulations
to determine how accurately this reduced-parameter-space model for $\xi$
reproduces the observed recoil velocities and find that
$\xi\approx145^\circ$, independent of either $q$ or $\Delta^z$.

The paper is organized as follows, in Sec.~\ref{sec:techniques}
we
review the numerical techniques used for the evolution of
the black-hole binaries and the analysis of the physical
quantities extracted at their horizons. 
In Sec.~\ref{sec:pn} we review the 
post-Newtonian dynamics of binary systems in order to
motivate our study of equivalent trajectories for unequal
mass, nonspinning and spinning holes. We focus on
four families of such configurations. In Sec.~\ref{sec:ID} we give
the  initial data parameters for these families.
The results of the evolution of
those configurations are given in Sec.~\ref{sec:res}, where we
also introduce a novel analysis of the trajectories of
the punctures and of the waveform phase to model the
angle $\xi$ in our heuristic formula Eq.~(\ref{eq:empirical}).
 In Sec.~\ref{sec:gen} we analyze the generic configuration
that led us to discover the large recoil velocities produced
by the spin projection on the orbital plane of the binary.
Here we use  more refined tools to analyze
the individual hole spins and momenta near merger time, when
most of the recoil is generated. We end the paper with
a Discussion section pointing out the need for further
runs with higher accuracy to improve our first results,
and an Appendix including the post-Newtonian analysis
of the maximum recoil configuration.

\section{Techniques}
\label{sec:techniques}

We use the puncture approach~\cite{Brandt97b} along with the {\sc
TwoPunctures}~\cite{Ansorg:2004ds} thorn to compute initial data.  In
this approach the 3-metric on the initial slice has the form
$\gamma_{a b} = (\psi_{BL} + u)^4 \delta_{a b}$, where $\psi_{BL}$ is
the Brill-Lindquist conformal factor, $\delta_{ab}$ is the Euclidean
metric, and $u$ is (at least) $C^2$ on the punctures.  The
Brill-Lindquist conformal factor is given by
$
\psi_{BL} = 1 + \sum_{i=1}^n m_{i}^p / (2 |\vec r - \vec r_i|),
$
where $n$ is the total number of `punctures', $m_{i}^p$ is the mass
parameter of puncture $i$ ($m_{i}^p$ is {\em not} the horizon mass
associated with puncture $i$), and $\vec r_i$ is the coordinate location of
puncture $i$.  We evolve these black-hole-binary data-sets using the
{\sc LazEv}~\cite{Zlochower:2005bj} implementation of the moving
puncture approach~\cite{Campanelli:2005dd}.  In our version of the
moving puncture approach~\cite{Campanelli:2005dd,Baker:2005vv} we replace the
BSSN~\cite{Nakamura87,Shibata95, Baumgarte99} conformal exponent
$\phi$, which has logarithmic singularities at the punctures, with the
initially $C^4$ field $\chi = \exp(-4\phi)$.  This new variable, along
with the other BSSN variables, will remain finite provided that one
uses a suitable choice for the gauge. An alternative approach uses
standard finite differencing of $\phi$~\cite{Baker:2005vv}. 

We use the Carpet~\cite{Schnetter-etal-03b,carpet_web} mesh refinement
driver to provide a `moving boxes' style mesh refinement. In this
approach  refined grids of fixed size are arranged about the
coordinate centers of both holes.  The Carpet code then moves these
fine grids about the computational domain by following the
trajectories of the two black holes.

We obtain accurate, convergent waveforms and horizon parameters by
evolving this system in conjunction with a modified 1+log lapse and a
modified Gamma-driver shift
condition~\cite{Alcubierre02a,Campanelli:2005dd}, and an initial lapse
$\alpha(t=0) = 2/(1+\psi_{BL}^{4})$.
The lapse and shift are evolved with
\begin{subequations}
\label{eq:gauge}
\begin{eqnarray}
\partial_t - \beta^i \partial_i) \alpha &=& - 2 \alpha K\\
 \partial_t \beta^a &=& B^a \\
 \partial_t B^a &=& 3/4 \partial_t \tilde \Gamma^a - \eta B^a.
 \label{eq:Bdot}
\end{eqnarray}
\end{subequations}
These gauge conditions require careful treatment of $\chi$, the
inverse
of the three-metric conformal factor,
near the puncture in order for the system to remain
stable~\cite{Campanelli:2005dd,Campanelli:2006gf,Bruegmann:2006at}. In
Ref.~\cite{Gundlach:2006tw} it was
shown that this choice of gauge leads to a strongly hyperbolic
evolution system provided that the shift does not become too large.

We use {\sc AHFinderdirect}~\cite{Thornburg2003:AH-finding} to locate
apparent horizons.
We measure the magnitude of the horizon spin using the Isolated
Horizon algorithm detailed in~\cite{Dreyer02a}. This algorithm is
based on finding an approximate rotational Killing vector (i.e.\ an
approximate rotational symmetry) on the horizon, and given this
approximate Killing vector $\varphi^a$, the spin magnitude is
\begin{equation}\label{isolatedspin}
S_{[\varphi]} = \frac{1}{8\pi}\oint_{AH}(\varphi^aR^bK_{ab})d^2V
\end{equation}
where $K_{ab}$ is the extrinsic curvature of the 3D-slice, $d^2V$ is the
natural volume element intrinsic to the horizon, and $R^a$ is the
outward pointing unit vector normal to the horizon on the 3D-slice.
We measure the
direction of the spin by finding the coordinate line joining the poles
of this Killing vector field using the technique introduced
in~\cite{Campanelli:2006fy}.  Our algorithm for finding the poles of
the Killing vector field has an accuracy of $\sim 2^\circ$
(see~\cite{Campanelli:2006fy} for details).

We also use an alternative quasi-local measurement of the spin and
linear momentum of the individual black holes in the binary that is
based on the
coordinate rotation and translation vectors~\cite{Krishnan:2007pu}.
In this approach the spin components of the horizon are given by
\begin{equation}
  S_{[i]} = \frac{1}{8\pi}\oint_{AH} \phi^a_{[i]} R^b K_{ab} d^2V,
  \label{eq:coordspin}
\end{equation}
where 
 $\phi^i_{[\ell]} = \delta_{\ell j} \delta_{m k} r^m \epsilon^{i j k}$,
and $r^m = x^m - x_0^m$ is the coordinate displacement from the centroid
of the hole,
while the linear momentum is given by
\begin{equation}
  P_{[i]} = \frac{1}{8\pi}\oint_{AH} \xi^a_{[i]} R^b (K_{ab} - K \gamma_{ab}) d^2V,
  \label{eq:coordmom}
\end{equation}
where 
 $\xi^i_{[\ell]} = \delta^i_\ell$.

We measure radiated energy, linear momentum, and angular momentum, in
terms of $\psi_4$, using the formulae provided in
Refs.~\cite{Campanelli99,Lousto:2007mh}. However, rather than using
the full $\psi_4$ we decompose it into $\ell$ and $m$ modes and solve
for the radiated linear momentum, dropping terms with $\ell \geq 5$. The
formulae in Refs.~\cite{Campanelli99,Lousto:2007mh} are valid at $r=\infty$. We obtain highly
accurate values for these quantities by solving for them on spheres of
finite radius (typically $r/M=25, 30, 35, 40$), fitting the results to
a polynomial dependence in $l=1/r$, and extrapolating to $l=0$. We
perform fits based on a linear and quadratic dependence on $l$, and
take the final values to be the average of these two extrapolations
with the differences being the extrapolation error.

\begin{figure}
\includegraphics[width=3.5in]{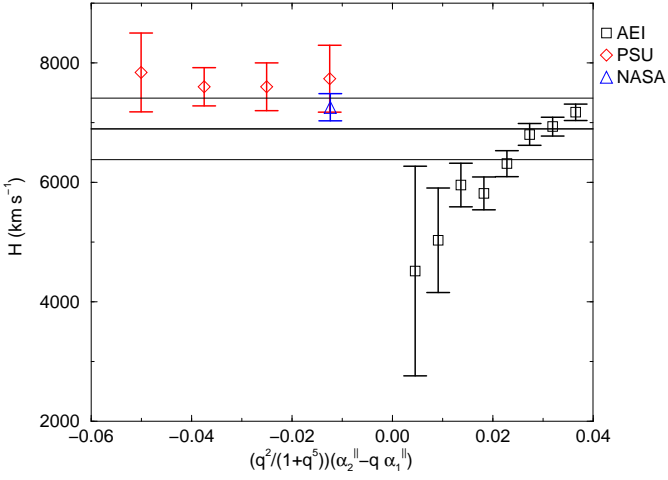}
\caption{The value of $H$ calculated by inverting Eq.~(\ref{eq:vperp})
as determined from simulations by the AEI, PSU, and
NASA/GSFC groups. The thick line is the weighted average and the thin
lines are the expected uncertainty in the average.}
\label{fig:H2}
\end{figure}

We obtain a new determination of $H$ in Eq.~(\ref{eq:vperp})
using results from simulations performed by the
NASA/GSFC~\cite{Baker:2007gi}, PSU~\cite{Herrmann:2007ac}, and
AEI/LSU~\cite{Pollney:2007ss} groups. The simulations performed by
these groups include runs with $q=1$, and thus provide an accurate
measurement of $v_\perp$. We calculate $H$ for each simulation
(via $H= v_\perp  (\alpha_2^\| - \alpha_1^\|) (1+q)^5/q^2$) and take
the weighted average $\avg{H}\pm \delta\avg{H}$, where
\begin{eqnarray}
\avg{X^n} &=& \sum_i {X_i}^n w_i, \nonumber \\
w_i &=& \frac{(\delta X_i)^{-2}}{\sum_j (\delta X_i)^{-2}},\nonumber \\
\delta \avg{X} &=& \sqrt{\avg{X^2} - \avg{X}^2},
\label{eq:avg}
\end{eqnarray}
$X$ is the quantity to be averaged, $n$ is some specified
power, and
$\delta X_i$ is the uncertainty in a particular measurement
of $X$. Note that we weight $H$ and $H^2$ by the same $w_i$.
We find $\avg{H} = (6895 \pm 513)\ \KMS$.
Figure~\ref{fig:H2} shows the values of $H$ obtained from each simulation
as well as the average value of $H$.
We can see that based
on the AEI/LSU data, which take into account the initial recoil
at the beginning of the full numerical simulations, one could
fit linear corrections to $H$. However, the deviations from $H={\rm const}$
are only significant near $D=q^2/(1+q)^5(\alpha_2^\| - q \alpha_1^\|)=0$, when
the spin-induced recoil is small (and hence the relative error in
the spin-induced recoil is large). The absolute differences between
the predicted and measured
recoil velocities for the AEI/LSU results are within $20\ \KMS$
when we take $H=6895\ \KMS$.

\section{Post-Newtonian analysis}
\label{sec:pn}
In order to compare results from the recoil due to unequal masses and
those due to spin effects as well, we will study systems with similar
orbital trajectories. Since the radiated momentum due to unequal masses is
a function of the orbital acceleration, these systems will all exhibit
very similar unequal-mass contributions to the net recoil, which allows
us to isolate the spin-induced contributions to the recoil.
To generate families of binaries with similar trajectories we use
the formulae for the leading order post-Newtonian accelerations and
choose configurations that minimize the effects of the spins on the
trajectories, but have non-negligible spin contributions
to the net recoil.

The relative one-body accelerations can be written as~\cite{Kidder:1995zr}
\begin{equation}
{\vec a} = {\vec a}_N + {\vec a}_{PN} + {\vec a}_{2PN}+ {\vec a}_{RR}
+ {\vec a}_{SO} + {\vec a}_{SS}, 
\label{accel}
\end{equation}
with
\newpage
\begin{widetext}
\begin{subequations}
\begin{equation}\label{eq:aN}
{\vec a}_N = - {m \over r^2} {\hat n} ,
\end{equation}
\begin{equation}\label{eq:aPN}
{\vec a}_{PN}=  - {m \over r^2} \left\{  {\hat n} \left[
(1+3\nu)v^2 - 2(2+\nu){m \over r} - {3 \over 2} \nu \dot r^2 \right]
-2(2-\nu) \dot r {\vec v} \right\} ,
\end{equation}
\begin{eqnarray}\label{eq:a2PN}
{\vec a}_{2PN} = - {m \over r^2} \biggl\{ && {\hat n} \biggl[ {3 \over 4}
(12+29\nu) ( {m \over r} )^2
+ \nu(3-4\nu)v^4 + {15 \over 8} \nu(1-3\nu)\dot r^4
\nonumber \\ && \mbox{}
- {3 \over 2} \nu(3-4\nu)v^2 \dot r^2
- {1 \over 2} \nu(13-4\nu) {m \over r} v^2
- (2+25\nu+2\nu^2) {m \over r} \dot r^2 \biggr]
\nonumber \\ && \mbox{}
- {1 \over 2} \dot r {\vec v}
\left[ \nu(15+4\nu)v^2 - (4+41\nu+8\nu^2)
{m \over r} -3\nu(3+2\nu) \dot r^2 \right] \biggr\} ,
\end{eqnarray}
\begin{equation}\label{eq:aRR}
{\vec a}_{RR} = {8 \over 5} \nu {m^2 \over r^3} \left\{ \dot r {\hat n}
\left[ 18v^2 + {2 \over 3} {m \over r} -25 \dot r^2 \right] - {\vec v}
\left[ 6v^2 - 2
{m \over r} -15 \dot r^2 \right] \right\} ,
\end{equation}
\begin{equation}\label{eq:aSO}
{\vec a}_{SO} = {1 \over r^3} \left\{ 6 {\hat n} [( {\hat n} \times
{\vec v} ) {\cdot} (2{\vec S} + {\delta m \over m}{\vec \Delta} )]
- [ {\vec v} \times (7
{\vec S}+3{\delta m \over m}{\vec \Delta})]
+ 3 \dot r [ {\hat n} \times
(3{\vec S} + {\delta m \over m}{\vec \Delta} )] \right\} ,
\end{equation}
\begin{equation}\label{eq:aSS}
{\vec a}_{SS} = - {3 \over \mu r^4} \biggl\{ {\hat n} ({\vec S_1 \cdot
\vec S_2}) + {\vec S_1} ({\hat n \cdot \vec S_2}) + {\vec S_2} ({\hat n \cdot
\vec S_1}) - 5 {\hat n} ({\hat n \cdot \vec S_1})({\hat n \cdot \vec S_2})
\biggr\} ,
\end{equation}
\end{subequations}
\end{widetext} 
where ${\vec x} \equiv {\vec x_1}-{\vec x_2}$, ${\vec v}={d{\vec x}/dt}$,
${\hat n}\equiv{{\vec x}/r}$, $m=m_1+m_2$, $\mu \equiv m_1m_2/m$, $\nu \equiv
\mu /m$, $\delta m \equiv m_1 - m_2$, ${\vec S} \equiv {\vec S_1}+{\vec S_2}$,
and ${\vec \Delta} \equiv m({\vec S_2}/m_2 -{\vec S_1}/m_1)$, and an overdot
denotes $d/dt$.

The first four terms in Eq.~(\ref{accel}) correspond to the Newtonian,
first-post-Newtonian (1PN), second-post-Newtonian, and radiation
reaction contributions to the equations of motion.
The last two terms in Eq.~(\ref{accel}) are the spin-orbit $(SO)$ and
spin-spin $(SS)$ contributions to the acceleration.

The radiated linear momentum due to the motion of the two holes
has the form~\cite{Kidder:1995zr}
\begin{eqnarray}
{\dot {\vec P}}_N = && - {8 \over 105} {\delta m \over m}
\nu^2 \left( {m \over r}
\right)^4 \biggl\{ \dot r { \hat n} \left[ 55v^2 -45\dot r^2 + 12 {m \over
r}
\right] \nonumber \\ && \mbox{}
+ {\vec v} \left[ 38\dot r^2 - 50v^2 - 8 {m \over r} \right] \biggr\},
\end{eqnarray}
plus higher post-Newtonian terms~\cite{wiseman:1992dv},
while the radiated linear momentum due to spin-orbit effects has the
form
\begin{eqnarray}
\label{eq:PSOdot}
{\dot {\vec P}}_{SO} = && - {8 \over 15} {\mu^2 m \over r^5}
\Bigl\{ 4 \dot r ({\vec v \times \vec\Delta})
- 2v^2 ({\hat n \times \vec\Delta}) \nonumber \\ && \mbox{}
- ({\hat n \times \vec{v}}) \left[ 3\dot r ({ \hat n \cdot \vec\Delta})
+ 2 ({\vec{v} \cdot \vec\Delta}) \right] \Bigr\}.
\end{eqnarray}
Note also that the spin-spin coupling does not
contribute to the radiated linear momentum to this order.

In order to best study and model how the final remnant
recoil velocity depends on the
mass ratio and spins, we will chose configurations with black holes
spinning along the orbital angular momentum. In this way the
orbital plane will not precess and we can write~\cite{Kidder:1995zr}
\begin{equation}\label{eq:S}
\vec S = \vec S_1 + \vec S_2 = S^z \hat z, \\
\end{equation}
and
\begin{equation}\label{eq:v}
\vec v = \dot{r}\hat{n}+r\omega\hat\lambda, \\
\end{equation}
where ${\vec L_N} \equiv \mu ({\vec x \times \vec{v}})$ is the
Newtonian orbital angular momentum, ${\hat \lambda} = {\hat L_N \times
\hat n}$ with ${\hat L_N} = {\vec L_N}/|{\vec L_N}|$, and $\omega=d\phi/dt$ is
defined as the orbital angular velocity.

Taking into account that the velocity remains in the orbital plane,
i.e. Eq.~(\ref{eq:v}), we find that the spin-orbit acceleration (\ref{eq:aSO})
is given by
\begin{equation}\label{eq:aSOorbit}
{\vec a}_{SO}^\perp = {1 \over r^3} \left\{ r\omega\left(5S^z+3{\delta m \over m}{\Delta}^z\right) {\hat n} -2\dot{r}S^z\hat\lambda\right\} ,
\end{equation}
and the radiated linear momentum is given by
\begin{eqnarray}\label{eq:PSOorbit}
{\dot {\vec P}}_{SO}^\perp = {16 \over 15} {\mu^2 m \Delta^z\over r^5}
\Bigl\{ (\dot r^2-r^2\omega^2) \hat\lambda
- 2\dot r r\omega\hat n \Bigr\},
\end{eqnarray}
and
\begin{eqnarray}\label{eq:PNorbit}
{\dot {\vec P}}_N = && - {8 \over 105} {\delta m \over m}
\nu^2 \left( {m \over r}
\right)^4 \biggl\{ \dot r { \hat n} \left[ 5r^2\omega^2 -2\dot r^2 + 4 {m \over
r}
\right] \nonumber \\ && \mbox{}
- r\omega{\hat\lambda} \left[50r^2\omega^2+ 12\dot r^2 + 8 {m \over r} \right] \biggr\}.
\end{eqnarray}
Note that if we take the scalar product of these two instantaneous radiated momenta
we obtain
\begin{eqnarray}\label{eq:PNPSO}
{\dot {\vec P}}_N \cdot {\dot {\vec P}}_{SO}^\perp / ({\dot {P}}_N {\dot {P}}_{SO}^\perp)
&&= \cos(\xi_{PN}^{inst}) \nonumber\\
&&=
-\omega f(r,\dot{r},\omega)/\sqrt{|g(r,\dot{r},\omega)|},
\end{eqnarray}
where 
$f=4r\dot{r}^4+(8m+24r^3\omega^2)\dot{r}^2
-r^2\omega^2(4m+25r^3\omega^2)$
and 
$g(r,\dot{r},\omega)=4\dot{r}^6-(16m-124r^3\omega^2-r)\dot{r}^4/r+
(16m^2+232r^3\omega^2m+1225r^6\omega^4+2r^4\omega^2)\dot{r}^2/r^2
+\omega^2(64m^2+800r^3\omega^2m+2500r^6\omega^4+r^4\omega^2)$.
The fact that the factor of $\Delta^z$
drops out
of Eq.~(\ref{eq:PNPSO}) suggests that $\xi$ (which is the angle between the
cumulative radiated linear momenta) will depend only weakly on the spins
through their affects on the orbital motion. Binaries with similar orbital
trajectories should therefore have similar values for
$\xi$. Note that $\xi$ may still be a strong function of trajectory
and $q$.

We will now turn to the question of identifying a subset of physical
parameters of the binary that produce similar trajectories for
unequal-mass, non-spinning and unequal-mass, spinning binaries in
order to compare their recoil velocities and extract the spin
contribution to the total recoil.

\subsection{similar radial trajectories}

An analysis of how $\xi$ depends on configuration is greatly simplified
if the trajectories of the spinning binaries are similar to
the trajectory for a non-spinning binary with the same mass ratio.
In order to accomplish this, we use the post-Newtonian expression
for the spin-orbit induced acceleration Eq.~(\ref{eq:aSOorbit}), and
choose configuration that minimize its effect.

The radial component of the spin-orbit induced acceleration will vanish if
$5S^z+3{\delta m \over m}{\Delta}^z=0$. This leads to the
condition
\begin{equation}\label{eq:F=0}
F=(3q+2)+(3+2q)\tilde{\alpha}=0,
\end{equation}
where $\tilde{\alpha}=q\alpha_1/\alpha_2$ can take any positive or negative value.
However, if we consider the algebraic average over the range $0\leq q \leq 1$ 
at fixed $F$ we find 
\begin{equation}
\avg{\tilde{\alpha}}=\frac12\left[\tilde{\alpha}(q=0)+\tilde{\alpha}(q=1)\right]
=\frac{4}{15}F-\frac{5}{6},
\end{equation}
and that $\tilde{\alpha} = \avg{\tilde{\alpha}}$ when $q=3/8$ (independent of
$F$).

We will thus study configurations with this mass
ratio (which also produces a nearly maximum recoil velocity of 
$\approx175\ \KMS$ for non-spinning unequal mass black hole binaries~\cite{Gonzalez:2006md}).

Hence the first family of black-hole-binary configurations that we will study
is given by the
choice
\begin{equation}\label{eq:F0q38}
F=0,\quad q=3/8,
\end{equation}
thus
\begin{equation}\label{eq:9/20}
\alpha_2/\alpha_1=-q(3+2q)/(2+3q)=-9/20.
\end{equation}
The total spin of the binary will in general be non-vanishing with
\begin{equation}\label{eq:SF0}
S^z/m^2=(\alpha_2+q^2\alpha_1)/(1+q)^2=4\alpha_2/11.
\end{equation}

\subsection{similar tangential trajectories}

We can also choose a configuration where the tangential component of the
acceleration due to the spin-orbit coupling vanishes,
i.e.
\begin{equation}\label{eq:S=0}
S^z=S^z_1+S^z_2=0.
\end{equation}
This translates into the condition
\begin{equation}\label{eq:9/64}
\alpha_2/\alpha_1=-q^2=-9/64
\end{equation}
when $q=3/8$.
Note that now, the radial acceleration, as parametrized by $F$, is non vanishing
\begin{equation}\label{eq:55/8}
F=(3q+2)+(3+2q)q\alpha_1/\alpha_2=-55/8.
\end{equation}
Thus, for $q\neq1$, we cannot make both the radial and
tangential components of the spin-orbit acceleration vanish at the same time
by a simple choice of physical parameters of the binary.

\section{Initial Configurations}
\label{sec:ID}

We choose quasi-circular initial configurations with mass ratio
$q=m_1/m_2=3/8$ from four
families of parameters that we will denote by Q, F, S, and A. The Q-series has
initially non-spinning holes, the F-series has $F=0$ (See Eq.~(\ref{eq:F=0}));
hence zero PN spin-orbit-induced radial acceleration,
the S-series has total spin $\vec S=0$; hence zero PN spin-orbit-induced
tangential acceleration, and the A-series has neither $F=0$
nor $S=0$; hence both spin-obit-induced accelerations are non-vanishing.
 The puncture masses were fixed by requiring that the total ADM mass of
the system be 1 and that the mass ratio of the horizon masses be 3/8. The initial
data parameters for these configurations are given in Tables~\ref{table:ID1} and \ref{table:ID2}.
 We obtained initial data parameters by choosing spin and linear
momenta consistent with 3PN quasi-circular orbits for binaries with mass ratio
$q=3/8$ and then solve for the Bowen-York ansatz for the initial 3-metric and extrinsic curvature.
This method was pioneered by the Lazarus project ~\cite{Baker:2002qf} 
(See Fig.~35 there), and then
used in the rest of the breakthrough papers~\cite{Campanelli:2006uy,
Campanelli:2006fg,Campanelli:2006fy,Campanelli:2007cg,Lousto:2007mh,
Krishnan:2007pu} by the authors (in Ref.~\cite{Campanelli:2007ew}
we used the PN expressions for the radial component of the momentum
as well).

\begin{widetext}

\begin{table}
\caption{Initial data parameters for quasi-circular orbits with orbital frequency
$\omega/M=0.05$. All sets have mass ratio $q=m_{1}^H/m_{2}^H=3/8$. The `F' series has
$\alpha=\alpha_2/\alpha_1=-9/20$ (hence $F=q\alpha_1/\alpha_2(2q+3)+3q+2=0$), and the `S' series
has $\vec S = \vec S_1 + \vec S_2 = 0$.
The punctures are located along the
$x$-axis with momenta $\vec P_1  = (0, P,0)$ and $\vec P_2 = (0,-P,0)$, and spins $\vec S_i$ along
the $z$-axis. $m_{i}^p$ are the puncture masses, $m_{i}^H$ are the horizon masses.}
\begin{ruledtabular}
\begin{tabular}{llllllll}
Config     & $\rm Q_{38}$  & $\rm F_{+0.2}$ & $\rm F_{-0.2}$   & $\rm F_{+0.4}$  & $\rm F_{-0.4}$  & $\rm S_{+0.64}$  & $\rm S_{-0.64}$    \\
\hline
$x_1/M$     &-4.7455652&-4.6889329  &-4.8008847  &-4.6310312  &-4.8548401  &-4.5310235 &-4.9561392\\
$x_2/M$     &1.7604572 &1.8161037   &1.7042740   &1.8711650   &1.6475993 &1.897592 &1.6168224\\
$S_1^z/M^2$ &0.0000000&0.015219622  &-0.015222140&0.030437161 &-0.030447242 &0.048726127 &-0.048689700\\
$S_2^z/M^2$ &0.0000000&-0.048702791 &0.048710847 &-0.097398914&0.097431175 &-0.048726127 &0.048689700\\
$P/M$       &0.10682112&0.10707929  &0.10656747  &0.10734244  &0.10631792  &0.10676349 &0.10692958\\
$L^z/M^2$   &0.69498063&0.6965546816 &0.6932382744&0.6979616178&0.6913258658 &0.6863415524 &0.7028440196\\
$J/M^2$     &0.69498063&0.6630715129 &0.7267269819&0.6309998643&0.7583097987 &0.6863415524 &0.7028440196\\
$m_{1}^p/M$&0.257487827988&0.25319314 &0.253279647&0.239665153&0.239816706&0.205915971&0.206131153\\
$m_{2}^p/M$&0.718534207968&0.71621170 &0.716211394&0.709030409&0.70903488&0.715832591&0.715746409\\
$m_{1}^H/M$ &0.27582974&0.27577886 &0.275791869&0.275757065&0.27577578&0.2756959&0.27558121\\
$m_{2}^H/M$ &0.73541100&0.73541402 &0.735444371&0.735334919&0.735402505&0.7351861&0.734888095\\
$\alpha_1^z$&0.000&0.20012582 &-0.20013825&0.4002982516&-0.400383662&0.6411766&-0.64119084\\
$\alpha_2^z$&0.000&-0.090053523&0.0900619119&-0.180130779&0.18015874&-0.090153&0.09015883\\
$M_{\rm ADM}/M $&1.00001&1.00001   &0.999997  &1.00001    &0.999997 &1.00001 & 0.999991\\

\end{tabular} \label{table:ID1} 
\end{ruledtabular} 
\end{table} 
\end{widetext} 

\begin{table}
\caption{
Initial data parameters for quasi-circular orbits with orbital frequency
$\omega/M=0.05$. All sets have mass ratio $q=m_{1}^H/m_{2}^H=3/8$. The punctures are located along the
$x$-axis with momenta $\vec P_1 = (0, P, 0)$ and $P_2 = (0,-P,0)$, and spins $\vec S_i$ along
the $z$-axis. $m_{i}^p$ are the puncture masses, $m_{i}^H$ are the horizon masses.
 In this series neither $F$ nor $S$ vanishes.}
\begin{ruledtabular}
\begin{tabular}{lllllll}
Config     & $\rm A_{+0.9}$ & $\rm A_{-0.9}$  \\
\hline
$x_1/M$    &-4.5443438&-4.8662563\\
$x_2/M$     &1.573114&1.9275192\\
$S_1^z/M^2$ &0.0000000&0.0000000\\
$S_2^z/M^2$ &0.48873779&-0.48581609\\
$P/M$    &0.10276465&0.11089309\\
$L^z/M^2$   &0.6286584770&0.7533827395\\
$J/M^2$     &1.117396265&0.2675666537\\
$m_{1}^p/M$ &0.2545666&0.2545806\\
$m_{2}^p/M$ &0.2822299&0.284150275\\
$m_{1}^H/M$ &0.2733564&0.2726292\\
$m_{2}^H/M$ &0.728824&0.7270093\\
$\alpha_1^z$&0.0000000&0.00000\\
$\alpha_2^z$&0.920196524 &-0.9192121\\
$M_{\rm ADM}/M $ &1.000000 &0.999991\\
\end{tabular} \label{table:ID2} 
\end{ruledtabular} 
\end{table} 

\section{Results}
\label{sec:res}

We evolved all configurations given in Tables~\ref{table:ID1}~and~\ref{table:ID2}
using 10 levels of refinement with a finest resolution of $h=M/80$ and
outer boundaries at $320M$ except configuration $\rm A_{+0.9}$, where we
used an additional coarse level to push the outer boundaries to $640M$.  In all
cases, except where noted otherwise, we set the free Gamma-driver parameter
in Eq.~(\ref{eq:Bdot})
to $\eta=6/M$~\cite{Alcubierre02a,Campanelli:2005dd}.

In a generic simulation both the unequal mass and spin components of
the recoil are functions of the trajectory. To single out each
individual effect we perform runs chosen to follow similar
trajectories.  In order to compare recoil velocity directions between
these runs we need to rotate each system so that the final plunge
(where most of the recoil is generated) occurs along the same
direction.  We do this in two ways. First, as demonstrated in
Fig.~\ref{fig:rot_fig_a2}, we plot the puncture trajectory difference
$\vec r = \vec r_1 - \vec r_2$ (where $\vec r_i (t)$ is the coordinate
location of puncture $i$ at time $t$) for each case and rotate the
trajectories by an angle $\Phi_{\rm track}$ so that they all line up
with the $\rm Q_{38}$ trajectory during the late inspiral and merger
phases. Note that by taking the differences between trajectories we
remove effects due to the wobble motion of the center of mass. Second,
we measure the phase of the dominant $(\ell=2,m=2)$ mode of $\psi_4$
at the point of peak amplitude and take half the phase difference
between each case and $\rm Q_{38}$ (a rotation of $\phi$ about the
$z$-axis will introduce a phase difference of $-2\phi$ in the $m=2$
components of $\psi_4$). We denote this latter rotation angle by
$\Phi_{\psi_4}$. We get reasonable agreement between these two
measures of the rotation angle (See Table~\ref{table:rot_compare}).
This type of rotation may also be needed when comparing results from
different resolutions of the same configuration (i.e.\ when the phase
error, but not the amplitude error, is large). In
Table~\ref{table:q38-q38-comp} we give the components of the recoil
velocity for a set of $Q_{38}$ simulations with $\eta=2/M$. This value
of $\eta$ leads to a poorer effective resolution than for our standard
choice of $\eta=6/M$. Consequently there is a relatively large phase
error in the low resolution results.  After performing the rotation,
the recoil velocities agree to within errors.

Note that there is no rotation which will make the $A_{+0.9}$ or
$A_{-0.9}$ trajectories line up with the $Q_{38}$ trajectory. In these
cases the hangup-effect~\cite{Campanelli:2006uy} due to spin-obit
coupling significantly alters the trajectories (See
Fig.~\ref{fig:rot_fig_a9}).

Once we have found the correct rotation angle we obtain $\xi$ via
\begin{eqnarray}
  \tilde{\vec V}_{\rm recoil} &=& R[\Phi] \vec V_{\rm recoil}, \nonumber \\
  \tilde{\vec V}_{\rm spin} &=& \tilde {\vec V}_{\rm recoil} - \vec V_{Q38}, \nonumber\\
  \cos(\xi) &=& \tilde{\hat V}_{\rm spin}\cdot\hat V_{Q38},
  \label{eq:xi_from_v}
\end{eqnarray}
where $\vec V_{\rm recoil}$ is the measured recoil velocity,
$R[\Phi]$ rotates $\vec V_{\rm recoil}$ by an angle $\Phi$
in the $xy$ plane, and $\vec V_{Q38}$ is the recoil of the
$Q_{38}$ configuration. Note that when $\alpha^\|_2 - q \alpha^\|_1 < 0$
we need to replace $\xi$ by $\pi - \xi$ in
formula~(\ref{eq:xi_from_v}) since the coefficient $v_\perp$ 
in Eq.~(\ref{eq:empirical}) is negative. We calculate 
two different values of $\xi$, $\xi_{\rm track}$ and $\xi_{\psi_4}$, 
based on the rotation angles $\Phi_{\rm track}$ and ${\Phi_{\psi_4}}$
respectively. We obtain an additional measurement of $\xi$ by solving
for $\cos \xi $ using Eq.~(\ref{eq:empirical}) and the measured values
of the recoil magnitude.
We denote
this latter measurement of $\xi$, which is unaffected by rotations,
by $\xi_{\rm Formula}$,
where
\begin{equation}\label{eq:xiformula}
  \xi_{\rm Formula} = \cos^{-1} \left[\frac{v^2-v_m(q)^2 -
   v_\perp(q,{\alpha_1}^\|, {\alpha_2}^\|)^2}
   {2 v_m(q)\,v_\perp(q,{\alpha_1}^\|, {\alpha_2}^\|)}\right],
\end{equation}
$v_m$ is given by Eq.~(\ref{eq:vm}), $v_\perp$ is given by Eq.~(\ref{eq:vperp}),
and $v$ is the measured magnitude of the recoil velocity.

\begin{table}
\caption{The rotation angle needed to align the trajectories of
each simulation with the $Q_{38}$ simulation as measured directly
from the orbital trajectories ($\Phi_{\rm track}$)
 and using the phase of the waveform at the
point of maximum amplitude ($\Phi_{\psi_4}$).
Note that $\Phi_{\psi_4}$ provides the
rotation angle modulo $180^\circ$. }
\begin{tabular}{lccc}
\hline\hline
Config   & $\Phi_{\rm track}$ & $\Phi_{\psi_4}$ &
              $|\Phi_{\rm track}-\Phi_{\psi_4}|$ \\
\hline
$\rm F_{+0.2}$  & $25^\circ$ & $34.5^\circ$ & $9.5^\circ$ \\
$\rm F_{-0.2}$  & $-28^\circ$ & $-35.5^\circ$ & $7.5^0$\\
$\rm F_{+0.4}$  & $56^\circ$ & $63.1^\circ$ & $7.1^\circ$ \\
$\rm F_{-0.4}$  & $-44^\circ$ & $-40.0^\circ$ & $4.0^\circ$ \\
$\rm S_{+0.64}$ & $5^\circ$ & $9.7^\circ$ & $4.7^\circ$\\
$\rm S_{-0.64}$ & $56^\circ$ & $44.6^\circ$ & $ 11.4^\circ$\\
$\rm A_{+0.9}$  & $***$ & $12.3^\circ$ & $***$\\
$\rm A_{-0.9}$  & $***$ & $-15.9^\circ$ & $***$\\
\hline\hline
\end{tabular} \label{table:rot_compare} 
\end{table} 

\begin{table}
\caption{Results of the recoil velocity for the $\rm Q_{38}$ configuration
with $\eta=2/M$ at two different resolutions. After correcting
for the phase error, equivalent to a rotation, the two recoils agree. Here `$\rm R_{track}$' denotes
the velocity after rotating by the angle $\Phi_{\rm track}$ and
`$R_{\psi_4}$' denotes
the velocity after rotating by the angle $\Phi_{\psi_4}$.}
\begin{tabular}{lcccc}
\hline\hline
$h$   & $\Phi_{\rm track}$ & $\Phi_{\psi_4}$ & $V_{x}$ & $V_{y}$\\
\hline
$M/80$ & $34^\circ$ & $36.5^\circ$ & $-163\pm12$&$-46\pm11$\\
$M/80$ $R_{\psi_4}$ & *** & *** & $-103\pm12$&$-134\pm11$\\
$M/80$ $\rm R_{track}$ & *** & *** & $-109\pm12$&$-129\pm11$\\
$M/100$ & $0$ & $0$ & $-109\pm14$&$-133\pm12$\\
\hline\hline
\end{tabular} \label{table:q38-q38-comp} 
\end{table}

We summarize the results of our simulations in Tables~\ref{table:resultsI}
and~\ref{table:resultsII}. All configuration, with the exception of the
`A' series, have radiated energies in the range 
$E_{\rm rad}/M = 0.021\pm0.002$ and radiated angular momenta in the
range $J_{\rm rad}/M^2 = 0.15\pm0.01$, which is consistent with these
trajectories being essentially the same for all configurations
(See Fig.~\ref{fig:rot_fig_a2}).

\begin{widetext}

\begin{figure}
\caption{The trajectory differences $\vec r = \vec r_1 - \vec r_2$ for 
the `F' and `S' series rotated so that the late-inspiral
matches the $\rm Q_{38}$
trajectory. The plots show the rotation angle $\Phi_{\rm track}$}

\includegraphics[width=2.7in]{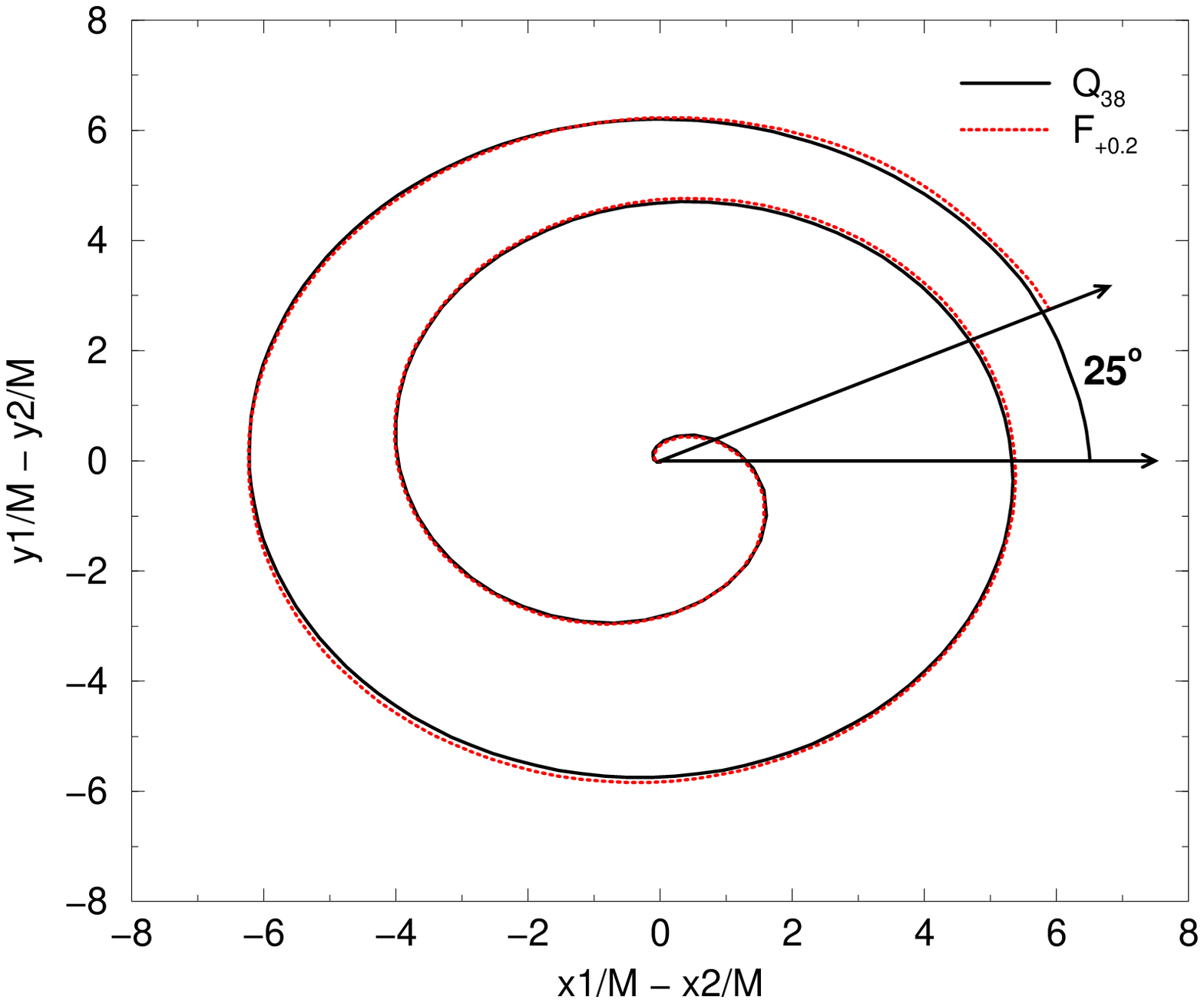}
\label{fig:rot_fig_a2}
\includegraphics[width=2.7in]{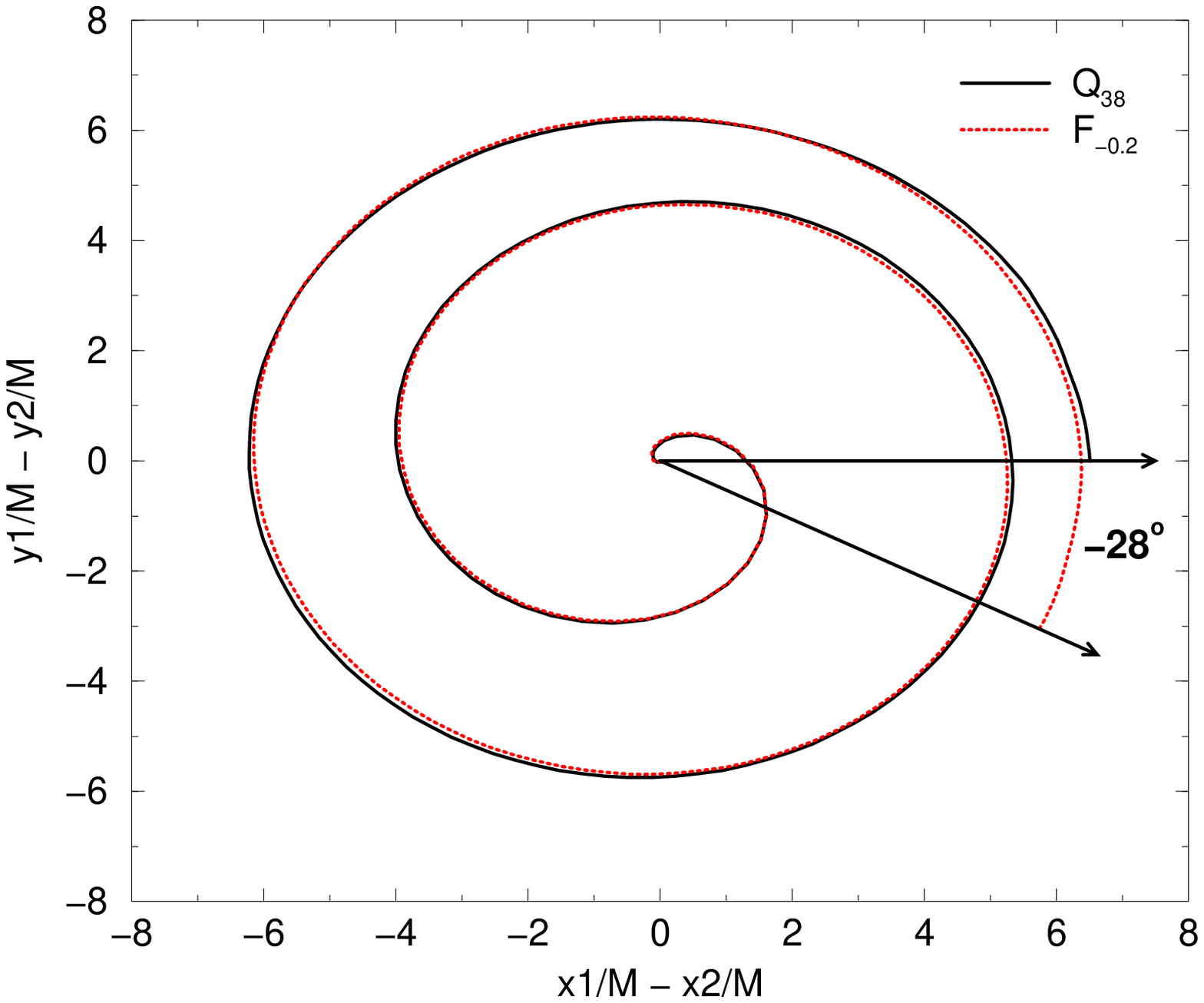}

\includegraphics[width=2.7in]{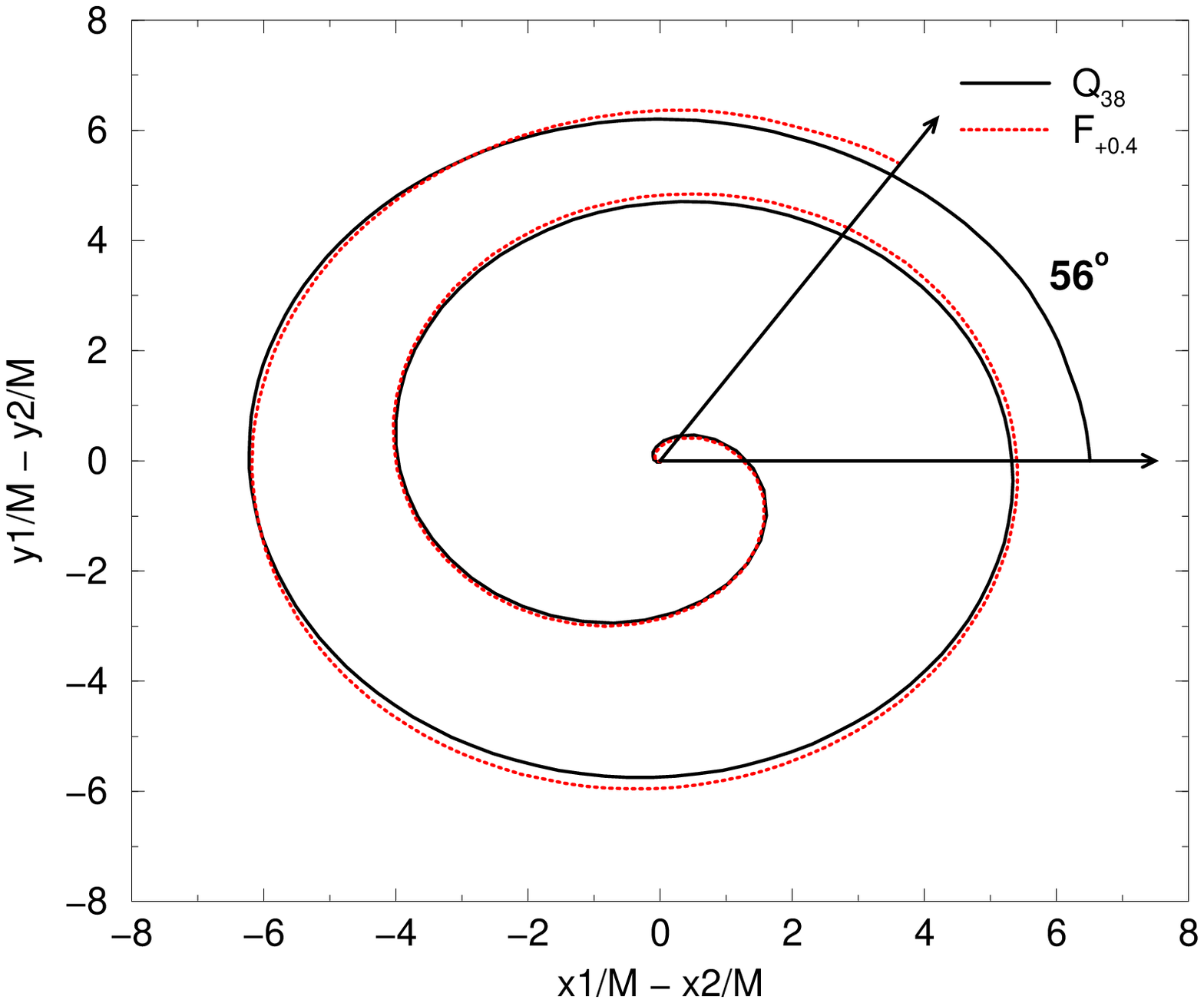}
\includegraphics[width=2.7in]{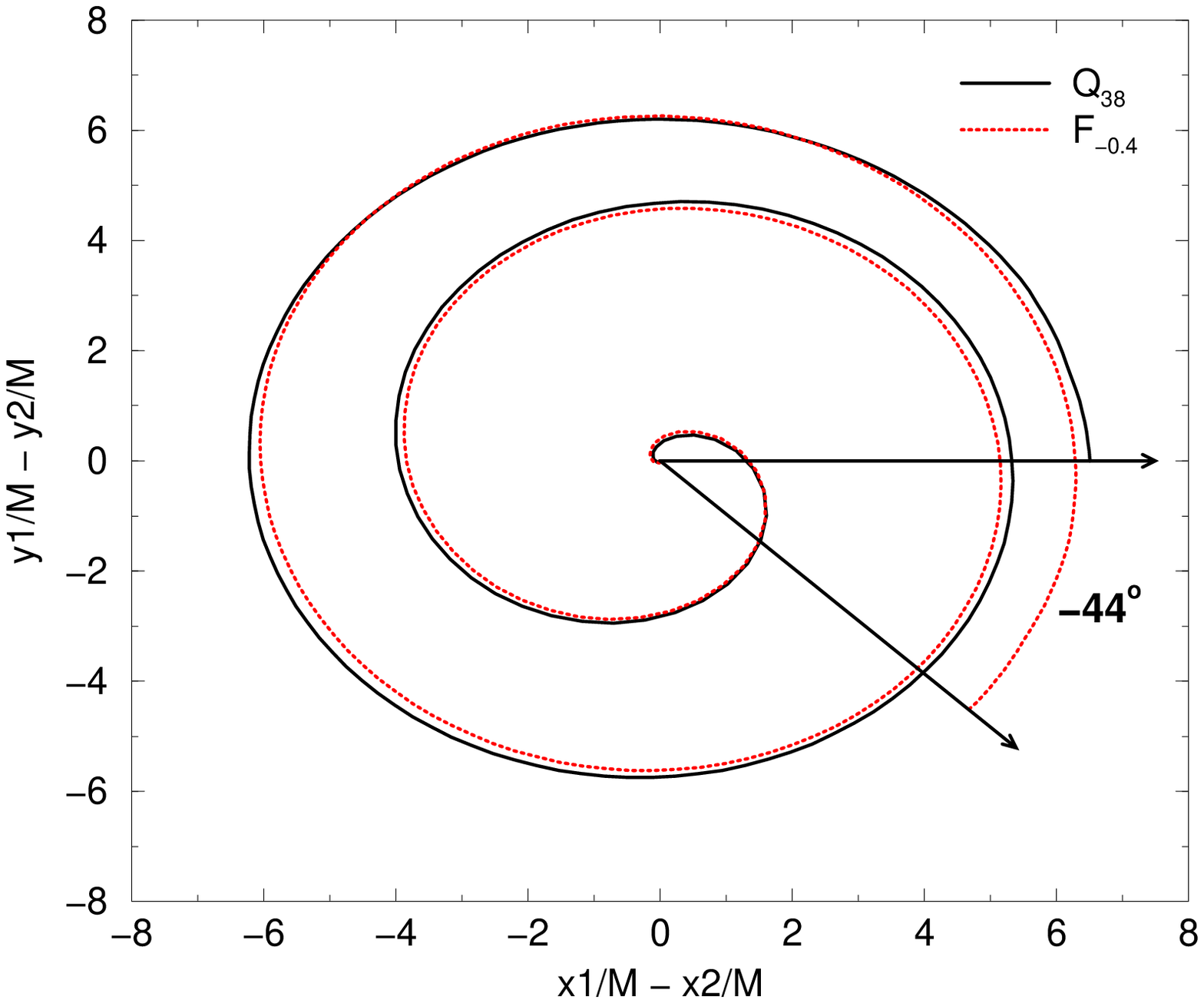}

\includegraphics[width=2.7in]{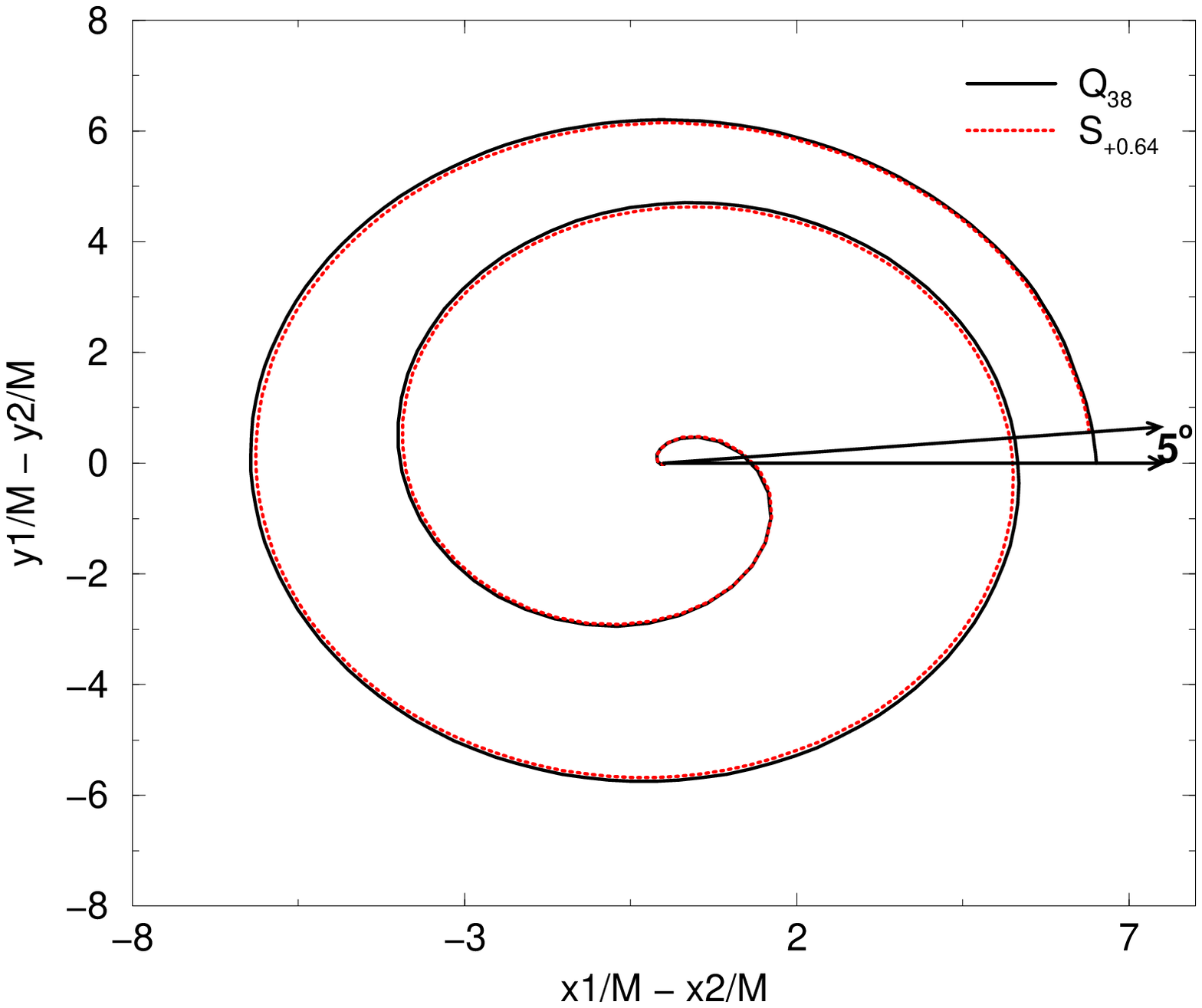}
\includegraphics[width=2.7in]{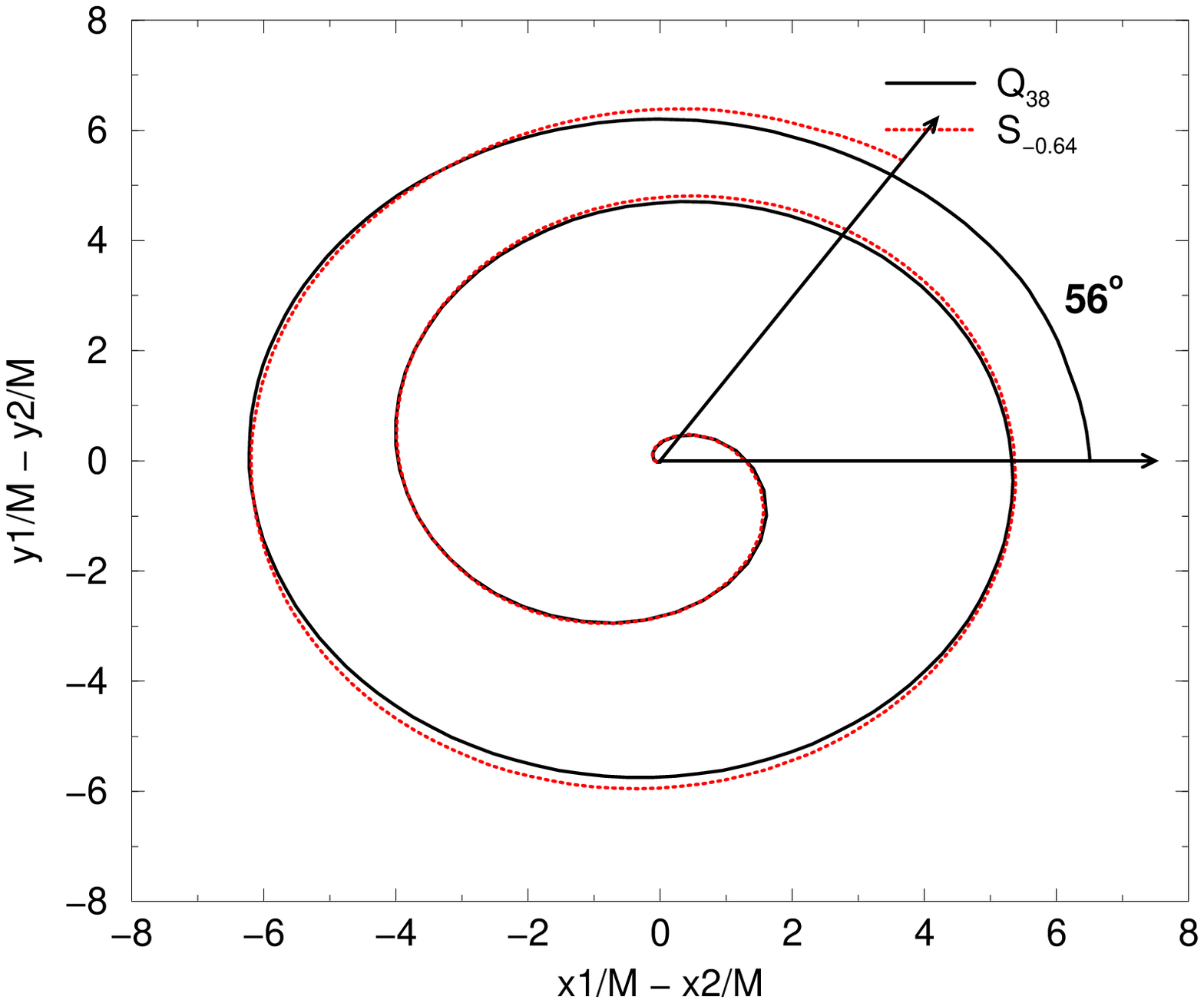}
\end{figure}

\begin{figure}
\caption{The trajectory differences $\vec r = \vec r_1 - \vec r_2$ for 
the `A' series, as well as  $\rm Q_{38}$. Note that there is no angle
$\Phi_{\rm track}$ that will make the late-time trajectories
overlap. Here the spin-orbit hang-up effect changes
the orbital trajectory significantly.}
\includegraphics[width=2.7in]{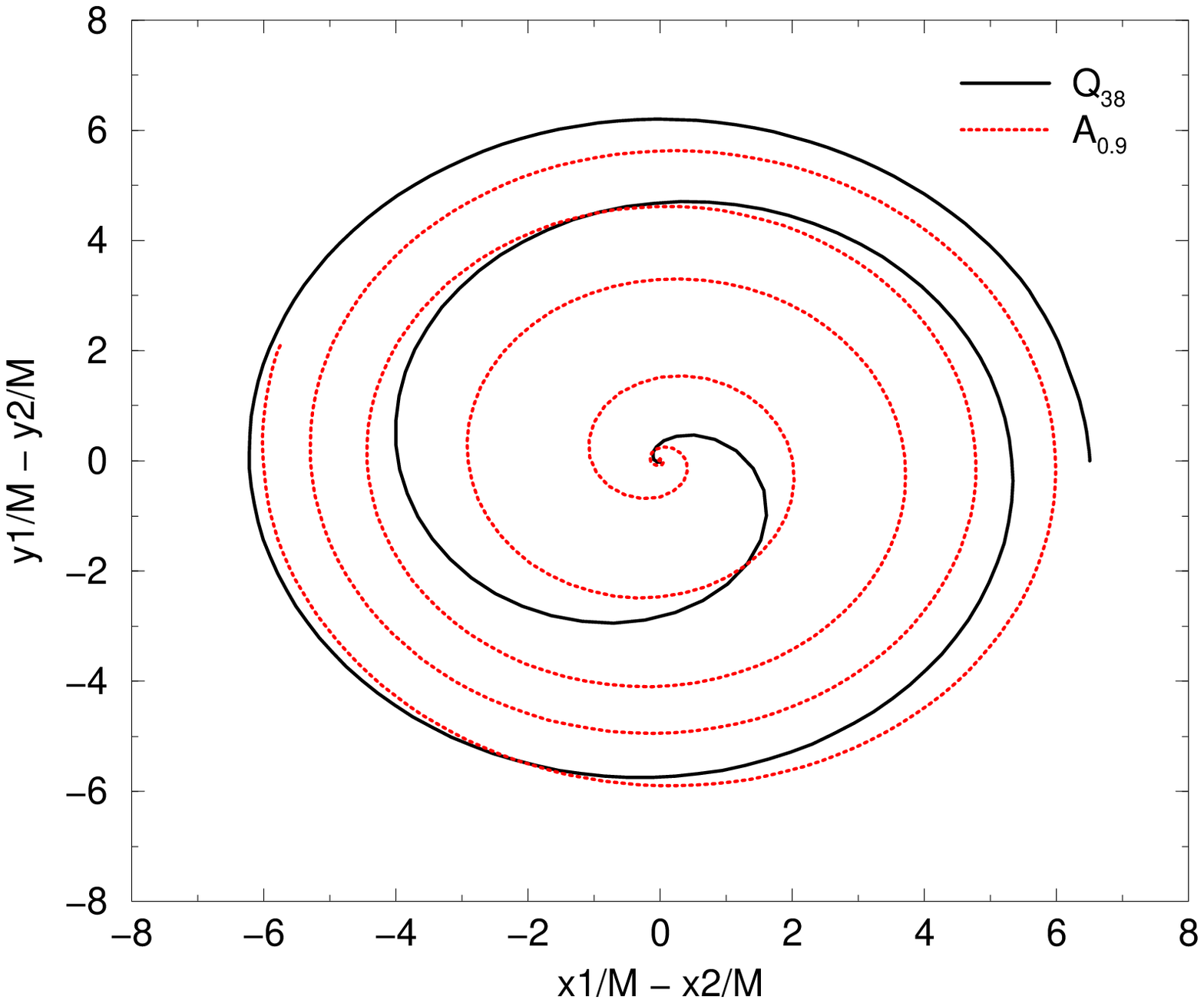}
\label{fig:rot_fig_a9}
\includegraphics[width=2.7in]{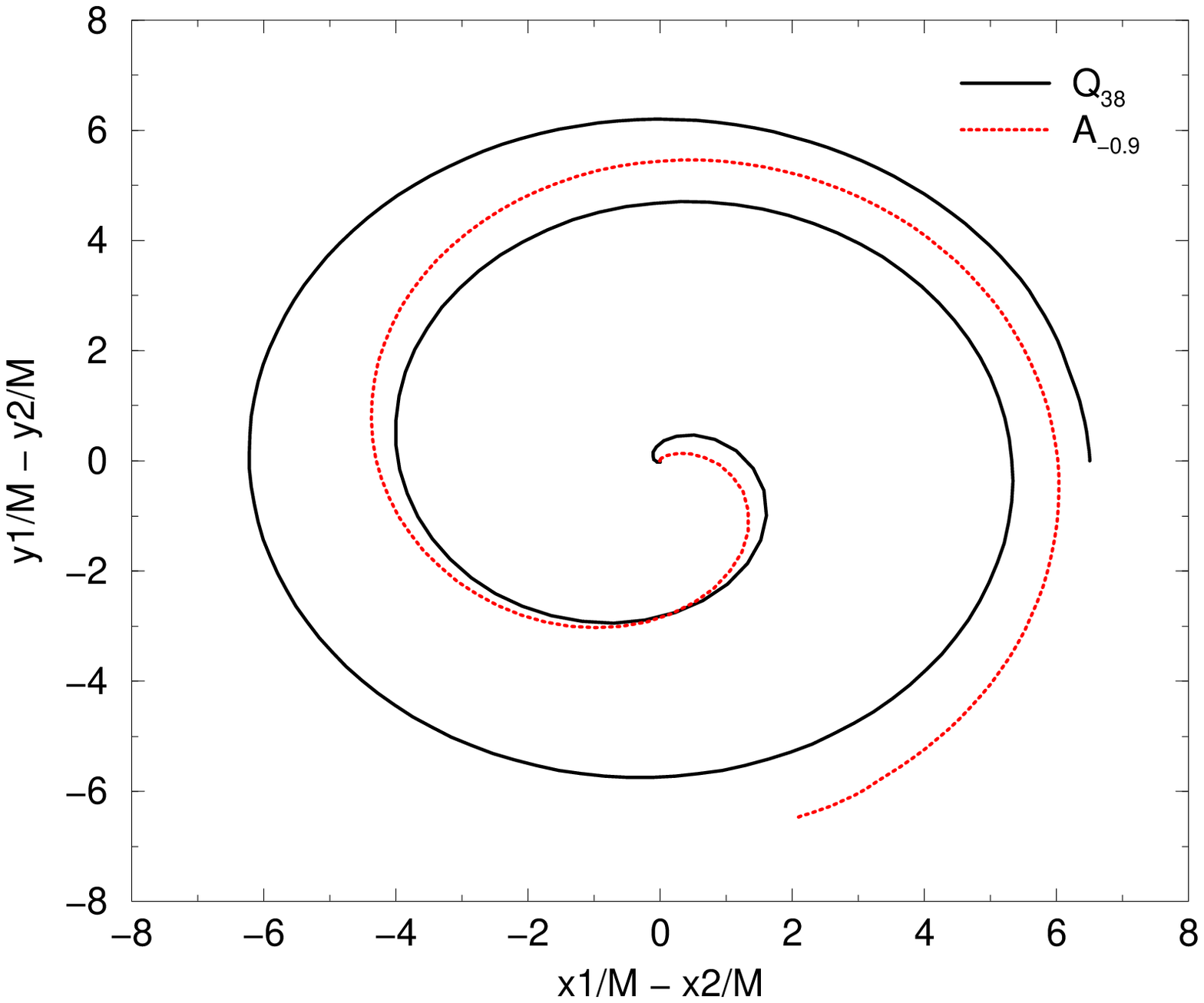}
\end{figure}

\begin{table}
\caption{The recoil velocities (prior to any rotation),
radiated energy and angular momentum, 
 and $\xi$ for the `Q' and `F' series. $\xi_{\rm track}$ is calculated
using $\Phi_{\rm track}$ and Eq.~(\ref{eq:xi_from_v}), $\xi_{\psi_4}$
is calculated
using $\Phi_{\psi_4}$ and Eq.~(\ref{eq:xi_from_v}), $\xi_{\rm Formula}$
is calculated from the given recoil magnitude using Eq.~(\ref{eq:xiformula}).
$|\vec V^{\rm pred}_{\rm track}|$, $|\vec V^{\rm pred}_{\psi_4}|$, and
$|\vec V^{\rm pred}_{\rm avg}|$  are
 the recoil velocities  as predicted by 
Eq.~(\ref{eq:empirical}) with $\xi = \xi_{\rm track}$, $\xi=\xi_{\psi_4}$,
and $\xi = \avg{\xi}$ respectively.
}
\begin{tabular}{lccccc}
\hline\hline
Config 	    	& $\rm Q_{38}$  & $\rm F_{+0.2}$  & $\rm F_{-0.2}$  & $\rm F_{+0.4}$  & $\rm F_{-0.4}$    \\
\hline
$E_{\rm rad}/M$		&$0.0210\pm0.0003$&$0.0202\pm0.0003$&$0.0219\pm0.0004$&$0.0193\pm0.0002$&$0.0228\pm0.0004$   \\
$J_{\rm rad}/M^2$	&$0.1503\pm0.0030$&$0.1471\pm0.0005$&$0.1576\pm0.0015$&$0.1399\pm0.0016$&$0.1625\pm0.0010$   \\
$V^x[\KMS]$	&$-94\pm11$&$-177\pm10$&$-15\pm14$&$-223\pm12$&$15\pm14$	\\
$V^y[\KMS]$	&$-141\pm5$&$-85\pm12$&$-155\pm5$&$33\pm18$&$-127\pm4$	\\
$|\vec V|[\KMS]$&$169.5\pm7.4$	&$196.4\pm10.4$	&$155.7\pm7.2$	&$225.4\pm12.2$	&$127.9\pm4.3$	\\
\hline
$\xi_{\rm track}$[deg]	&$0^\circ$   &$(143\pm31)^\circ$ &$(178\pm73)^\circ$	&$(147\pm20)^\circ$	& $(169\pm21)^\circ$	\\
$\xi_{\psi_4}$[deg]	&$0^\circ$   &$(154\pm43)^\circ$ &$(127\pm41)^\circ$	&$(173\pm25)^\circ$	& $(179\pm21)^\circ$	\\
$\xi_{\rm Formula}$[deg]	&$0^\circ$   &$(127\pm26)^\circ$ &$(131\pm15)^\circ$	&$(134\pm20)^\circ$	& $(144\pm6)^\circ$	\\
\hline
$|\vec V^{\rm pred}_{\rm track}|[\KMS]$&175	&$202\pm9$	&$142\pm3$	&$232\pm10$	&$112\pm9$	\\
$|\vec V^{\rm pred}_{\psi_4}|[\KMS]$&175	&$205\pm9$	&$158\pm21$	&$240\pm5$	&$110\pm5$	\\
$|\vec V^{\rm pred}_{\rm avg}|[\KMS]$&175	&$203\pm3$	&$150\pm4$	&$231\pm5$	&$127\pm8$	\\
\hline\hline
\end{tabular} \label{table:resultsI} 
\end{table} 
\end{widetext} 

\begin{widetext}

\begin{table}
\caption{The recoil velocities (prior to any rotation), radiated energy and angular momentum, and
$\xi$ for the `S' and `A' series. Note that although we report the
calculated values for $\xi$ based on $\Phi_{\psi_4}$ for the
`A' series, here $\xi$ is not well defined because
the unequal mass component of the recoil is not
given by the $Q_{38}$ recoil. $\xi_{\rm track}$ is calculated
using $\Phi_{\rm track}$ and Eq.~(\ref{eq:xi_from_v}), $\xi_{\psi_4}$
is calculated
using $\Phi_{\psi_4}$ and Eq.~(\ref{eq:xi_from_v}), $\xi_{\rm Formula}$
is calculated from the given recoil magnitude using Eq.~(\ref{eq:xiformula}).
$|\vec V^{\rm pred}_{\rm track}|$, $|\vec V^{\rm pred}_{\psi_4}|$, and
$|\vec V^{\rm pred}_{\rm avg}|$  are
 the recoil velocities  as predicted by 
Eq.~(\ref{eq:empirical}) with $\xi = \xi_{\rm track}$, $\xi=\xi_{\psi_4}$,
and $\xi = \avg{\xi}$ respectively.
}
\begin{tabular}{lcccc}
\hline\hline
Config 	    	& $\rm S_{0+0.64}$  & $\rm S_{-0.64}$  & $\rm A_{+0.9}$  & $\rm A_{-0.9}$    \\
\hline
$E_{\rm rad}/M$		&0.0209$\pm 0.0003$&$0.0203\pm0.0002$&$0.050668\pm0.000974$&$0.01274\pm0.00003$\\
$J_{\rm rad}/M^2$		&$0.152\pm0.0007$&$0.146\pm0.001$&$0.2999857\pm0.00708$&$0.092\pm0.001$\\
$V^x[\KMS]$	&$-122\pm18$&$-119\pm5$&$13\pm30$&$48\pm24$	\\
$V^y[\KMS]$	&$-181\pm15$&$31\pm4$&$-63\pm2$&$-340\pm8$	\\
$|\vec V|[\KMS]$&$218.3\pm16.0$	&$123.0\pm4.9$	&$64.1\pm5.9$	&$343.4\pm8.6$	\\
\hline
$\xi_{\rm track}$[deg]	&$(160\pm31)^\circ$   &$(148\pm11)^\circ$	&$***$	&$***$	\\
$\xi_{\psi_4}$[deg]	&$(142\pm28)^\circ$   &$(137\pm7)^\circ$	&$(158\pm7)^\circ$	&$(93\pm7)^\circ$	\\
$\xi_{\rm Formula}$[deg]	&$(124\pm22)^\circ$   &$(150\pm7)^\circ$	&$(159\pm2)^\circ$	&$(149\pm19)^\circ$	\\
\hline
$|\vec V^{\rm pred}_{\rm track}|[\KMS]$&$237\pm10$	&$125\pm10$	&$***$	&$***$\\
$|\vec V^{\rm pred}_{\psi_4}|[\KMS]$&$230\pm16$	&$135\pm7$	&$68\pm22$	&$259\pm18$\\
$|\vec V^{\rm pred}_{\rm avg}|[\KMS]$&$231\pm5$	&$127\pm8$	&$108\pm28$	&$340\pm9$\\
\hline\hline
\end{tabular} \label{table:resultsII} 
\end{table} 
\end{widetext}

We obtain weighted averages for $\xi$ for the `F' and `S' series
of $\avg{\xi_{\rm track}} = (152\pm9)^\circ$, $\avg{\xi_{\psi_4}} = (143\pm14)^\circ$,
and $\avg{\xi_{\rm Formula}} = (144\pm7)^\circ$, where we use Eq.~(\ref{eq:avg})
to obtain the weighted average and uncertainty.
These weighted averages are consistent with the measured values of
$\xi$. The weighted average over all three measurements of $\xi$ is
$\avg{\xi} = (145\pm10)^\circ$.
Interestingly, $\avg{\xi}$ provides an accurate prediction for the
recoil velocity of the $A_{-0.9}$ configuration. This result is unexpected
because the recoil due
to unequal masses is a function of the mass ratio and the trajectories
(i.e.\ the accelerations of the masses over time). For the `F' and
`S' configuration the trajectories are very similar to $Q_{38}$, and
hence the unequal mass components of the recoil are expected to be
 very similar to $Q_{38}$. However,
the spin-orbit coupling induced hangup effect in both $A_{+0.9}$ and
$A_{-0.9}$ greatly affects the trajectories (See Fig.~\ref{fig:rot_fig_a9}),
as well as the radiated energy and angular 
momenta. If we consider the radiated linear momentum
averaged over an orbit, then we see that the slower the inspiral (i.e.\
the closer to a closed orbit), the
smaller the average recoil. Hence we expect that $A_{+0.9}$ will
have a smaller unequal-mass recoil than $Q_{38}$, while $A_{-0.9}$
will have a larger one. To
quantify how much the orbits close we take the average of
$\vec r = \vec r_1 - \vec r_2$ over the trajectory from the beginning of
each simulation until $|\vec r| \sim 0.1$. 
The
resulting averages $|\avg{\vec r}|$ for the `Q', `F', `S', and
`A' families are given in Table~\ref{table:ravg}. The mean and
standard deviation of $|\avg{\vec r}|$ for the `Q', `F', and `S' configurations
is $|\avg{\vec r}| = 0.865\pm0.070$. The $A_{+0.9}$ and $A_{-0.9}$ configuration
lie $7.1\sigma$ and $7.6\sigma$ below and above this mean respectively,
while the results for the other configurations lie within $1.4\sigma$ of
the mean.

\begin{table}
\caption{The average value $|\avg{\vec r}|$ of the trajectories for
each configuration. The larger the value of  $|\avg{\vec r}|$ the
slower the inspiral.
}
\begin{tabular}{llllll}
\hline\hline
Config  & $|\avg{\vec r}|$ & Config & $|\avg{\vec r}|$ & Config  & $|\avg{\vec r}|$\\
$Q_{38}$ & $0.858303$ & $F_{+0.2}$ & $0.902234$ & $F_{-0.2}$ & $0.7982157$\\
$F_{+0.4}$ & $0.936172$ & $F_{-0.4}$ & $0.76745$ & $S_{+0.64}$ & $0.845197$\\
$S_{-0.64}$ & $0.95333$ & $A_{+0.9}$ & $0.365654$ & $A_{-0.9}$ & $1.39869$\\
\hline\hline
\end{tabular} \label{table:ravg} 
\end{table}

\begin{figure}
\includegraphics[width=3.5in]{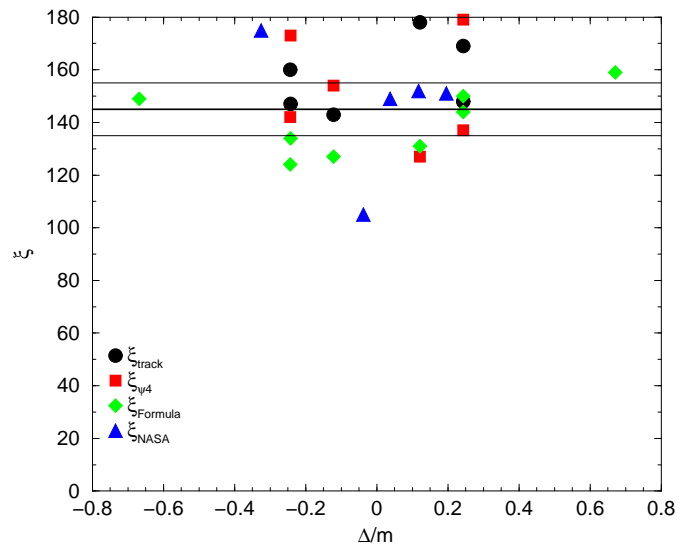}
\caption{$\xi$ versus $\Delta/m = S_2/m_2 - S_1/m_1$ as calculated in this work
for a mass ratio $q=3/8$ and
  from the data published by the NASA/GSFC group  for a mass ratio $q=2/3$
provided in Ref.~\cite{Baker:2007gi}.
  We plot $\xi_{\rm track}$, $\xi_{\psi_4}$, and $\xi_{\rm Formula}$ for the
  `F' and `S' configurations and $\xi_{\rm Formula}$ for `A' configurations.
  The thick horizontal line and the two thin horizontal line show the average value 
$\avg{\xi}$ and its uncertainty (as calculated in this work from our
simulations).
The data displays significant scatter, but appears to be
consistent with $\xi = {\rm const}$.
  }
\label{fig:XivsDelta}
\end{figure}

As seen in Fig.~\ref{fig:XivsDelta} the angle $\xi$ appears, at least
qualitatively, to be independent of $\Delta$. This is in agreement
with our post-Newtonian analysis in Eq.~(\ref{eq:PNPSO}).  It is also
consistent with our intuition that similar trajectories imply similar
angles between the unequal-mass and spin contributions
to the recoil, and it seems that the small differences in the trajectories
produce some scatter on the values, but this is apparently
mostly due to the numerical error generated during the simulations. It
would be interesting to use this value of $\xi$ to  model the recoil
velocity distribution in galaxies.

\section{generic evolution reanalyzed}
\label{sec:gen}

In light of our new understanding about the modeling of the recoil
velocity, we re-examine our original generic binary configuration,
which we denote by SP6. The SP6 configuration has a mass ratio of
$q=1/2$ with the larger hole having specific-spin $a/m = 0.885$
and spin pointing $45^\circ$ below the orbital plane, and the smaller
hole having negligible spin. We also evolved a similar configuration, which
we will denote by SP6R, that is identical to SP6, but with the spin
rotated by $90^\circ$ about the $z$-axis. We evolved both
configurations using the same grid structure as in the previous
section, but used $\eta=2/M$ rather than $6/M$. This choice of smaller
$\eta$ has the effect of reducing the effective resolution, but makes
calculations of the quasi-local linear momentum and spin direction more 
accurate (See Ref.~\cite{Krishnan:2007pu}) by reducing coordinate
distortions. The initial data parameters for the two configurations
are given in Table~\ref{table:id_sp6}. The drop in effective resolution when
reducing $\eta$ from $6/M$ to $2/M$ is significant. In our simulations we
found that a $Q_{38}$, $\eta=2/M$ run with central resolution of $M/100$ had
a slightly
larger waveform phase error than an equivalent $M/80$ resolution 
run with $\eta=6/M$, while an $M/80$ run with $\eta=2/M$ displayed a
significant phase error. We have found in general that, with our choice
of gauge, the coordinate dependent measurements, such as spin and linear
momentum direction, become more accurate as $\eta$ is reduced (and
$h\to 0$). However, if $\eta$ is too small ($\eta \lesssim 1/M$), the runs
may become unstable. Similarly, if $\eta$ is too large ($\eta \gtrsim 10/M$),
then grid stretching effects can cause the remnant horizon to continuously
grow, eventually leading to an unacceptable loss in accuracy at late-times.
We have found that a value of $\eta= 6/M$ provides both very high accuracy
in the computed waveform at modest resolutions, while keeping the 
remnant horizon size nearly fixed at late-times.

\begin{table}
\caption{Initial data parameters for the SP6 and SP6R
configurations. $m_p$ is the puncture mass parameter of the two holes.
SP6 has spins $\vec S_1 = (0, S, -S)$ and $\vec S_2 = (0,0,0)$,
momenta $\vec P = \pm (P_r, P_\perp, 0)$, puncture positions $\vec x_1 = (x_+, d, d)$ and
$\vec x_2 = (x_-, d, d)$, masses $m_1$ and $m_2$, and
$M_{\rm ADM}/M = 1.00000\pm0.00001$. SP6R has the same
parameters as SP6 with the exception that
$\vec S_1 = (-S, 0, -S)$.
}
\begin{tabular}{llllll}
\hline\hline
$m_p/M$ & $0.3185$  & $d/M$   & $0.0012817$ & $m_1/M$  & $0.6680$\\
$x_+/M$ & $2.68773$ & $P_r/M$ & $-0.0013947$ & $m_2/M$  & $0.3355$\\
$x_-/M$ & $-5.20295$ & $P_\perp/M$ &  $0.10695$ & $S/M^2$ & $0.27941$\\
\hline\hline
\end{tabular} \label{table:id_sp6}
\end{table}

We measure a net recoil of $V_{\rm recoil} = 375\pm18\ \KMS$ and
 $V_{\rm recoil} = 848\pm20\ \KMS$ for SP6 and SP6R respectively.

\begin{figure}
\includegraphics[width=3.5in]{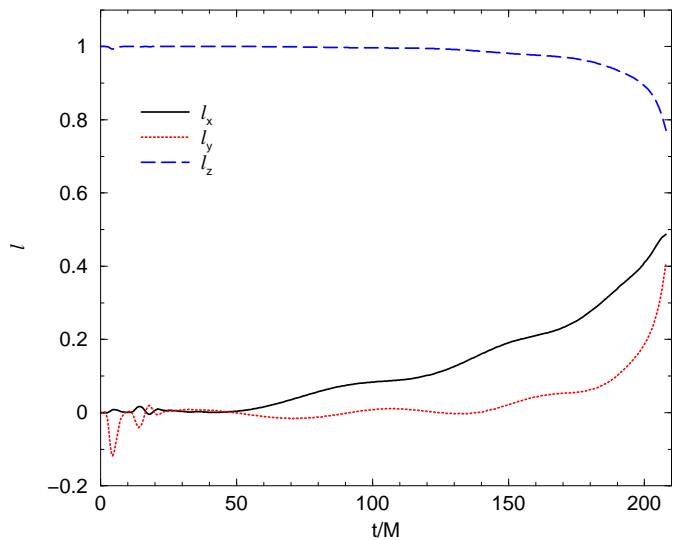}
\caption{The normalized orbital angular momentum vector
$\vec \ell = \vec L / |\vec L|$ versus time for the SP6 configuration up
to merger.
Note the rapid change in the direction at late times.}
\label{fig:sp6_orb_mom}
\end{figure}

\begin{figure}
\includegraphics[width=3.5in]{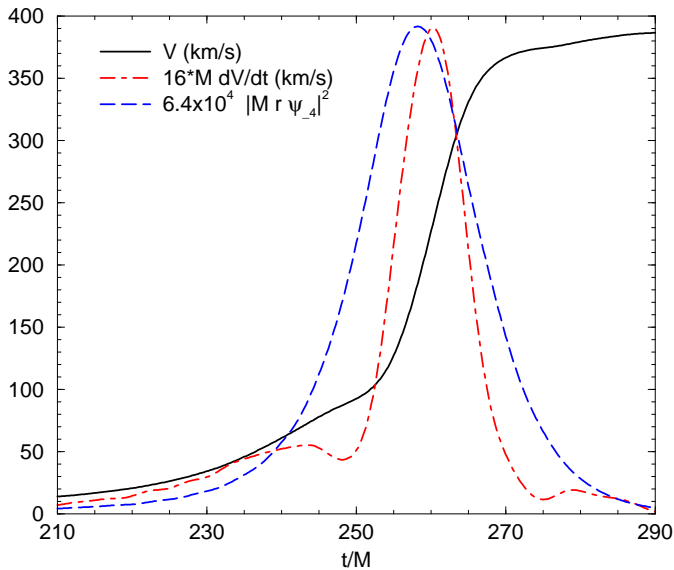}
\caption{The recoil speed ($V = |\vec V|$) for the SP6 configuration 
as measured from $\psi_4$
at $r=40M$ as
a function of time, as well as the time derivative of the recoil speed
($dV/dt = \hat V \cdot \dot{\vec V}$), and the
magnitude of $\psi_4$. Here the initial data burst is
excluded from the calculation. Note that peak in $dV/dt$ is located
between $t=250M$ and $t=270M$ and occurs about $2M$ latter than the
peak in $|\psi_4|$. A common horizon was first detected at  $t=207.4M$,
strongly suggesting that most of the recoil velocity is built up around
merger time (since the observer is at $r=40M$, features in the waveform 
at time $t=\tau$ originated
near the horizon(s) at time $t\sim\tau -40M$).}
\label{fig:sp6_kick_v_time_noburst}
\end{figure}

The analysis of the recoil in SP6 and SP6R is complicated by the fact
that the orbital plane precesses significantly during the merger.
Thus, we cannot associate the $xy$ components of the recoil with the
in-plane recoil (as was done tentatively in Ref.~\cite{Campanelli:2007ew}). In
order to measure the precession of the orbital plane we need an
accurate measurement of the orbital angular momentum. Here we use the
approximate formula
\begin{equation}
  \label{eq:coordang}
  \vec L_{\rm orbit} = \sum_i \vec r_i \times \vec P_i,
\end{equation}
where $\vec r_i$ is the coordinate location of puncture $i$ and
$\vec P_i$ is the quasi-local momentum~\cite{Krishnan:2007pu}, given
by Eq.~(\ref{eq:coordmom}), of black hole $i$.
In
Fig.~\ref{fig:sp6_orb_mom} we show the orbital angular momentum of the SP6
configuration versus time. Note the rapid change in direction near
merger (a common horizon was first detected at $t=207.4M$),
and as seen in Fig.~\ref{fig:sp6_kick_v_time_noburst},
most of the recoil is generated about $3M$ to $30M$ after merger 
(here we assume that waveform features seen at $t=\tau$ 
for an observer at $r=40M$ were generated by dynamics near the horizons at
$t\sim \tau - 40M$). This rapid change in direction has a strong
effect on the computed
recoil due to the $\cos \Theta$ and $\cos \xi$
 dependence of $v_{\rm recoil}$. That is, rapid physical changes in the orbital
plane and spin direction, lead to relatively large
errors in the direction (but not magnitude) of both the spin and
orbital angular momenta when the resolution is below some threshold. This in
turn, leads to relatively large errors in the measured recoil. Thus
it is not surprising that this new calculation of the recoil velocity for
SP6  is $100\ \KMS$ smaller than the value we reported
 in~\cite{Campanelli:2007ew} (note that we used a higher effective
resolution in~\cite{Campanelli:2007ew}, thus we expect those values
to be more accurate).
These large errors will not be observed in more symmetric binaries
where either the spin or angular momentum axes are fixed.

We can obtain an approximate measurement of $\alpha_\|$ and $\alpha_\perp$
using Eq.~(\ref{eq:coordang}) and the measured direction of the spin.
This estimation is only approximate due to the coordinate dependent nature
of both calculations. We find that for SP6, $\alpha_\|$ and $\alpha_\perp$
 vary little over the course of the run
with values at merger of $\alpha_\| = -0.62\pm0.03$ and $\alpha_\perp =
0.62 \pm0.03$ (which are within errors of the initial values).
 However, the SP6R configuration does show a definite
change in $\alpha$ over time, with merger values of 
$\alpha_\| = -0.69\pm0.03$ and $\alpha_\perp = 0.54\pm0.03$. 
We can use Eq.~(\ref{eq:empirical}) to give estimates for the predicted
recoil velocity if we make the following assumptions: (1) $\xi = \avg{\xi}$,
(2) $\Theta$ for SP6R is rotated by $\pi/2$ radians with respect to
SP6, and (3) $\Theta_0$ is the same for SP6 and SP6R. Given these assumptions
and the above range of the values for $\alpha_\|$ and $\alpha_\perp$, we
can perform a non-linear least-squares fit of the recoil magnitude for
SP6 and SP6R to obtain $\Theta_0$. The resulting predictions for the 
recoil magnitude are $V_{\rm SP6} = (500\pm60)\ \KMS$ and
$V_{\rm SP6R} = (1120\pm130)\ \KMS$. Both predictions are within $2 \sigma$ of
the actual measured values and have an absolute error of $32\%$.
If we fix $\alpha_\|$ and $\alpha_\perp$ to their average values and
vary our guess for $\xi$ over the range $(0, 360^\circ)$, we find that
the predicted values for $V_{\rm SP6}$ and $V_{\rm SP6R}$ lie in the
ranges $(462, 495)\KMS$ and $(1048, 1120)\KMS$ respectively.

The SP6 configuration demonstrated that the in-plane
component of the spin can be the dominant contribution to the recoil.
Given this observation, it becomes very important to accurately model
this recoil. In Appendix~\ref{sec:PNAppendix} we derive a post-Newtonian
model for the recoil produced by this in-plane component and show that it
predicts the $\cos \Theta$ dependence in our empirical formula.

\section{Discussion}

Interestingly, most of the recoil velocity imparted to the remnant is
generated at around merger time (more precisely, as seen in
Fig.~\ref{fig:sp6_kick_v_time_noburst}, within the first
few tens of $M$ after merger. See also Refs.~\cite{Baker:2006vn,
Campanelli:2007ew,Brugmann:2007zj}.),
a nonlinear regime where post-Newtonian
approximations
are not expected to work, but where the `Lazarus' approach
\cite{Baker:2000zh,Baker:2001nu,Baker:2001sf,Baker:2002qf,
Baker:2003ds,Campanelli:2005ia}
can be successfully applied~\cite{Campanelli:2004zw}.
 
Although an accurate modeling of $\xi$ is challenging, starting from
an ansatz that $\xi = \xi(q,\Delta)$, we have found that, for
quasi-circular orbits, $\xi$ is
qualitatively independent of either $\Delta$ or $q$ for $q=3/8$,
$q=2/3$ (based on the results of Ref.~\cite{Baker:2007gi}), and 
$q=1/2$ (based on SP6). Note that the $\xi$ that we measure is
consistent with a similar parameter introduced in
Ref.~\cite{Baker:2007gi}, where they found $\xi=147^\circ$ (in our
notation), based on a least-squares fit of the magnitude of the recoil
versus a simplified version of Eq.~(\ref{eq:empirical}).
We know from the results for headon collision (where $\xi = \pi/2$),
that $\xi$ is a function of eccentricity. However, for quasi-circular
orbits, it appears to vary only marginally with either $q$ or $\Delta$.
 Further long-term simulations with high-accuracy
(including extrapolations to $h\to0$ and $\eta\to0$) and 
further separated binaries  will be needed in
order to obtain a highly accurate model for $\xi$. 
In particular, the $\eta\to0$ limit will be important because
the recoil depends sensitively on the linear momenta and spin
directions of
the individual black holes near merger (where gauge effects are most
severe), and hence we need to take the $\eta\to0$ limit in order to
accurately measure $\vec \alpha$, $\vec L$, and $\Theta$.
Nevertheless, our
 simple formula holds with enough accuracy for astrophysical applications.
In particular we have seen that the determination of an average value for the
angle $\xi$ of $145^o$ seems to work not only for the 
$F$ and $S$ sequences, but also when we move off of these sequences
towards more generic binaries.  However, the formula should definitely 
be used with caution in an untested regime, especially when the trajectories
are significantly altered by spin-orbit effects.

\acknowledgments
We thank the referee for many helpful suggestions in improving
the text.  We gratefully acknowledge NSF for financial support
from grants PHY-0722315, PHY-0722703, PHY-0714388, PHY-653303.
Computational resources were provided by Lonestar cluster at TACC
and by NewHorizons at RIT.

\appendix
\section{Post-Newtonian modeling}\label{sec:PNAppendix}

Here we provide a brief post-Newtonian analysis of the configurations
that maximize the recoil velocity for spinning black holes.
The spin-orbit-coupling (SO) contribution to the radiated linear momentum
is given by Eq.~(\ref{eq:PSOdot}).

We will restrict our analysis to planar orbits.
Hence we have
\begin{equation}\label{eq:vagain}
\vec v = \dot{r}\hat{n}+r\omega\hat\lambda,
\end{equation}
where ${\hat \lambda} = {\hat L_N \times
\hat n}$, ${\hat L_N} = {\vec L_N}/|{\vec L_N}|$, $\omega$ is
the orbital angular velocity, and ${\vec L_N} \equiv \mu ({\vec x \times \vec{v}})$ is the
Newtonian orbital angular momentum.
We shall take  ${\hat L_N}\equiv \hat{z}$. Hence
\begin{equation}
\hat{\lambda}=\hat{z}\times\hat{n}\quad{\rm and}\quad
\hat{n}\times\hat{\lambda}=\hat{n}\times(\hat{z}\times\hat{n})=\hat{z}
\end{equation}

We observe that the third and fourth terms in Eq.~(\ref{eq:PSOdot})
only contribute to the recoil along the $z$-axis since
\begin{equation}
\hat{n}\times\vec{v}=r\omega\hat{z}
\end{equation}
This contribution to the recoil velocity might well be the leading one, hence, in order
to maximize the total recoil we seek to align, as much as possible, the first two
terms in Eq.~(\ref{eq:PSOdot}) with the $z$-axis. This is achieved by having 
the spin of the black holes lie in the orbital plane, i.e.
\begin{equation}
\vec\Delta=\Delta_n\hat{n}+\Delta_\lambda\hat{\lambda}.
\end{equation}
We then explicitly obtain the following products
\begin{eqnarray}
\vec{v}\times\Delta&=&(\dot{r}\Delta_\lambda-r\omega\Delta_n)\hat{z},\\
\hat{n}\times\Delta&=&\Delta_\lambda\hat{z},\\
\vec{v}\cdot\vec\Delta&=&\dot{r}\Delta_n+r\omega\Delta_\lambda.
\end{eqnarray}

Plugging this into Eq.~(\ref{eq:PSOdot})  we find
\begin{eqnarray}\label{eq:PSOdotZ}
{\dot {\vec P}}_{SO}^{\ \|}=- {8 \over 15} {\mu^2 m \over r^5}
\Bigl\{(2\dot{r}^2-4r^2\omega^2)\Delta_\lambda
-9\dot{r}r\omega\Delta_n
\Bigr\}\hat{z}.
\end{eqnarray}

This clearly displays the fact that the recoil will be maximized when $\Delta$
takes the maximum magnitude (equal mass and opposite maximally rotating black holes)
and varies sinusoidally with its projection along the line joining the holes.
Note that if we define the angle between $\hat{n}$ and $\vec\Delta$ as $\theta$ 
we can write the above equation as
\begin{eqnarray}\label{eq:PSOdottheta}
{\dot {\vec P}}_{SO}^{\ \|}&=&A(r)|\Delta|\cos\theta+B(r)|\Delta|\sin\theta\nonumber\\
&=&C(r)|\Delta|\cos(\theta-\theta_0(r)).
\end{eqnarray}
This $\cos \theta$ dependence in the recoil was the motivation for
proposing the now-verified $\cos \Theta$ dependence in our empirical
formula Eq.~(\ref{eq:vpar}) for the recoil.

Note that this analysis applies to the radiated linear
momentum flux. Hence we have assumed that the larger the radiated
linear momentum flux, the larger the total radiated linear momentum.

It is also interesting to see if the unexpectedly large magnitude of the
maximum out-of-plane recoil, compared to the in-plane recoil, can be
understood using the post-Newtonian expression for the radiated linear
momentum, i.e.\ Eqs.~(\ref{eq:PSOorbit})~and~(\ref{eq:PSOdotZ}) 
(See Ref.~\cite{Schnittman:2007ij} for a similar analysis). To do
this, we used the post-Newtonian formulae for the radiated linear
momentum along with the numerical trajectories for runs with the spins
in the plane and perpendicular to the plane. We found that the
post-Newtonian formulae predicted that the maximum out-of plane
recoil will be approximately twice (almost 9/4) as large, rather than
(the observed) $\approx8$ times as large, as the maximum in-plane
recoil. Thus we see that the magnitude of the out-of plane recoil
arises from nonlinear dynamics at merger not fully captured by the
post-Newtonian formalism.  One may then conclude that, while the
post-Newtonian approximation gives the correct dependence of the
recoil on the physical parameters, such as the scaling of the recoil
velocities with the components of the spins parallel and perpendicular
to the angular momentum, it is much less accurate when describing the
amplitude of the recoils.  Thus we find that post-Newtonian formalisms
provides the correct form for our semi-empirical
formula~(\ref{eq:empirical}), but does not provide accurate
measurements of the magnitudes of the constants in that formula.

\bibliographystyle{apsrev}
\bibliography{../../Lazarus/bibtex/references}

\end{document}